\documentclass[prb,aps,twocolumn]{revtex4-1}

\usepackage{amsmath}  
\usepackage{amsthm}  
\usepackage{amssymb}	
\usepackage{wasysym} 
\usepackage{graphicx}  
\usepackage[usenames,dvipsnames]{color}
\usepackage{array} 
\usepackage{paralist} 
\usepackage{amsfonts}
\usepackage{mathtools}
\usepackage{times}
\usepackage{braket}
\usepackage{slashed}
\usepackage{multirow}
\usepackage[colorlinks=true,linkcolor=blue,citecolor=blue,breaklinks]{hyperref}
\normalfont


\renewcommand{\vec}[1]{\ensuremath{\mathbf{#1}}} 
 
\newcommand{\abs}[1]{\left| #1 \right|} 
\newcommand{\avg}[1]{\left< #1 \right>} 
 
\let\baraccent=\= 
\renewcommand{\=}[1]{\stackrel{#1}{=}} 

\newcommand{\be}{\begin{equation}}
\newcommand{\ee}{\end{equation}}

\newcommand{\bea}{\begin{eqnarray}}
\newcommand{\eea}{\end{eqnarray}}
\newcommand{\beal}{\begin{align}}
\newcommand{\eeal}{\end{align}}

\usepackage{tikz}
\usetikzlibrary{shapes}

\tikzset{Boctagon/.style ={shape=regular polygon, regular polygon sides=8, draw, sharp corners}}
\tikzset{dodecagon/.style ={shape=regular polygon, regular polygon sides=12, draw, sharp corners}}

\newcommand{\Boctagon}[1]{\begin{tikzpicture}\node[Boctagon, scale  = #1]{};\end{tikzpicture}}
\newcommand{\dodecagon}[1]{\begin{tikzpicture}\node[dodecagon, scale = #1]{};\end{tikzpicture}}

\begin{document}

\title{Generic Field-Driven Phenomena in Kitaev Spin Liquids:\\ Canted Magnetism and Proximate Spin Liquid Physics}

\author{Ciar\'{a}n Hickey}
\email[E-mail: ]{chickey@thp.uni-koeln.de}
\affiliation{Institute for Theoretical Physics, University of Cologne, 50937 Cologne, Germany}
\author{Matthias Gohlke}
\affiliation{Theory of Quantum Matter Unit, Okinawa Institute of Science and
Technology Graduate University, Onna-son, Okinawa 904-0495, Japan}
\author{Christoph Berke}
\affiliation{Institute for Theoretical Physics, University of Cologne, 50937 Cologne, Germany}
\author{Simon Trebst}
\affiliation{Institute for Theoretical Physics, University of Cologne, 50937 Cologne, Germany}

\begin{abstract}
Topological spin liquids in two spatial dimensions are stable phases in the presence of a small magnetic field, 
but may give way to field-induced phenomena at intermediate field strengths. Sandwiched between the low-field 
spin liquid physics and the high-field spin-polarized phase, the exploration of magnetic phenomena in this intermediate
regime however often remains elusive to controlled analytical approaches.
Here we numerically study such intermediate-field magnetic phenomena for two representative Kitaev models 
(on the square-octagon and decorated honeycomb lattice) that exhibit
either Abelian or non-Abelian topological order in the low-field limit. Using a combination of  exact 
diagonalization and density matrix renormalization group techniques, as well as linear spin-wave theory, we
establish the generic features of Kitaev spin liquids in an external magnetic field. 
While ferromagnetic models typically exhibit a direct transition to the polarized state at a relatively low field strength, 
antiferromagnetic couplings not only substantially stabilizes the topological spin liquid phase, but generically lead to 
the emergence of a distinct field-induced intermediate regime, separated by a crossover from the high-field polarized regime. Our results suggest that, for most lattice geometries, this regime 
generically exhibits significant spin canting, antiferromagnetic spin-spin correlations, and an extended proximate spin liquid regime 
at finite temperatures. Notably, we identify a symmetry obstruction in the original honeycomb Kitaev model that prevents, at least for certain field directions, the formation of such canted magnetism without breaking symmetries -- consistent with the recent numerical observation of an extended gapless spin liquid in this case. 
\end{abstract}
\maketitle


\section{Introduction}

Quantum spin liquids (QSLs) are a fascinating example of the ``More is Different'' philosophy of modern condensed matter physics \cite{Anderson1972}, featuring fractionalized excitations and emergent gauge structures \cite{Savary2017quantum} that can only exist within the confines of a many-body system. Typically arising in systems which feature an element of frustration, thus hindering conventional magnetic ordering even at the lowest temperatures, they come in many different flavors, depending on the physical dimension and nature of the emergent excitations and underlying gauge structure. Though there are numerous material candidates, exhibiting various expected experimental signatures, it has proven difficult to unambiguously identify a QSL in nature \cite{Knolle2019FieldGuide}.    

\begin{figure}[b]  
\includegraphics[width=\columnwidth]{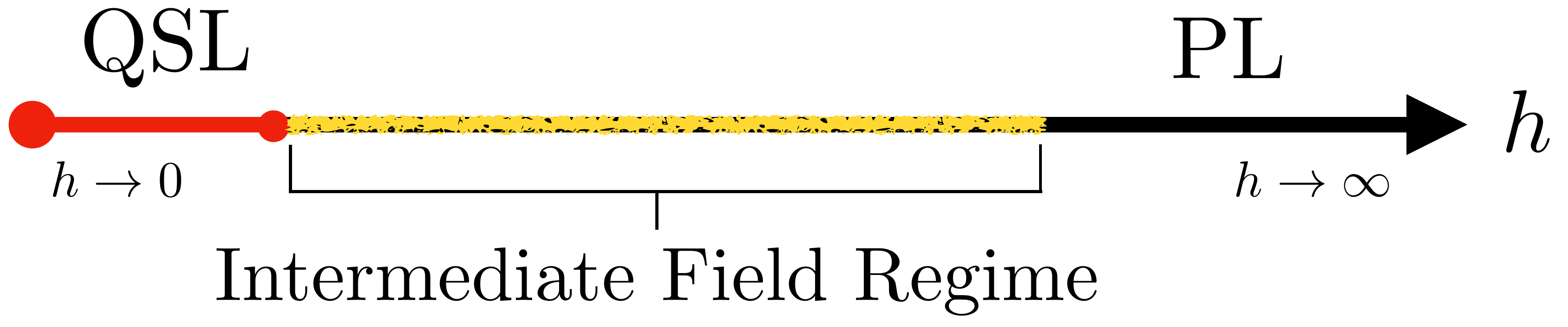}
\caption{\textbf{Generic phase diagram} of a gapped QSL in a magnetic field with a finite extent of the QSL at low fields, 
			a trivial polarized state (PL) at high fields, and a non-universal intermediate field regime.}
\label{fig:GenericPhasediagram}
\end{figure}

For a QSL, or indeed any magnetic phase of matter, a natural question to ask is what happens when an external magnetic field is applied? More precisely, if we have a Hamiltonian \[H=H_{\text{QSL}}-\sum_i\vec{h}\cdot \vec{S}_i\,,\] where the ground state of $H_{\text{QSL}}$ is a QSL and $\vec{h}$ points in some fixed direction, what is the resulting phase diagram as a function of the field magnitude $h=\abs{\vec{h}}$? 
The physics in the two extreme limits of such a phase diagram can be immediately deduced. (i) In the high-field limit, $h\rightarrow\infty$, 
the system should obviously be in a trivial polarized, or partially polarized, phase (depending on whether there is a conserved spin rotational symmetry about the field direction or not). In this limit the ground state is a simple product state, or approaches one as $h\rightarrow\infty$, and its excitations can be accurately described by linear spin-wave theory (LSWT). (ii) On the other hand, in the infinitesimal field limit, $h \rightarrow 0$, many QSLs remain stable up to some finite critical field strength $h_c$, their existence not relying on any symmetry requirements. Gapped QSLs are certainly stable, with the gap providing protection against infinitesimal perturbations. However there are some special cases of gapless QSLs that are believed to be unstable, even against an infinitesimal field, e.g.~they may harbor a fermionic spectrum with pairing instabilities that can be immediately triggered by a finite field \cite{Ran2009}. 
We are thus left with an (almost) generic phase diagram in which, in the high-field limit, there is a polarized state, and, in the low field limit, the QSL (in most cases) remains stable up to some finite critical field $h_c$. This leaves a wide model-dependent and non-universal region of ``intermediate fields'', between the low-field QSL and the high-field polarized state as depicted in Fig.~\ref{fig:GenericPhasediagram}, in which new physics may emerge. It is precisely examples of this region that we wish to explore in this manuscript.   

In order to study the physics at ``intermediate fields'' we should ideally choose a model in which the physics in the two limits, $h\rightarrow \infty$ and $h\rightarrow 0$, is already well-known. In this case we can keep track of what happens as we move toward the intermediate field regime, from both above and below. Unfortunately, though the physics of the $h\rightarrow \infty$ limit is relatively easy to describe, it is in general an extremely challenging task to accurately determine the physics in the $h\rightarrow 0$ limit, i.e.~to accurately describe the properties of $H_{\text{QSL}}$. In order to make progress we will specialize here to studying the Kitaev model \cite{Kitaev2006}, which consists of bond-dependent nearest-neighbor Ising interactions between spin-1/2 degrees of freedom on a tricoordinated lattice. The ground state of the model is known to be a QSL, which we refer to as the Kitaev spin liquid (KSL), allowing us to replace $H_{\text{QSL}}$ with $H_{\text{KSL}}$. 
The primary advantage of using the Kitaev model is that, owing to its special structure, it is exactly solvable, with its ground state, finite temperature and dynamical properties all well-known and signatures characterized \cite{Hermanns2018physics}. Furthermore it can be defined on a range of 2d and 3d tricoordinated lattices
\cite{yao-kivelson,Yang2007mosaic,Kells2011Kaleidoscope,Si2008anyonic,Mandal2009exactly,Hermanns2014quantum,Kimchi2014three,Hermanns2015weyl,Hermanns2015spin-peierls,Obrien2016classification,Yamada2017,Eschmann2020classification}, giving a whole family of models within which to explore. However, one must keep in mind that the model's unique properties are not necessarily common to all QSLs, or even all QSLs with a similar gauge structure. As a result one must always be cautious in extending any lessons learned from studies of the Kitaev model to other QSLs. 

We have now sharpened our initial, general yet unwieldy question of what happens to a QSL when an external field is applied to the more specific and manageable question of what happens to the KSL when an external field is applied? For example, what is the critical field $h_c$ for the KSL and how does it differ between different lattices? What is the field at which LSWT about the polarized phase begin to fail? Are there any new phases or phenomena that appear at intermediate fields? 
A remarkable example in this regard is the original honeycomb Kitaev model, for which it was recently demonstrated 
that a distinct gapless QSL appears at intermediate fields for antiferromagnetic (AFM) couplings 
by a variety of techniques \cite{zhu_robust_2018,gohlke_dynamical_2018,hickey_emergence_2019,liang_intermediate_2018,jiang_field_2018,patel_magnetic_2019,Berke2020}.
Does similar physics play out on other tricoordinated lattices? Is there any universal behavior to the KSL in a magnetic field across different lattices? 

To summarize our main conclusions, we will argue in this manuscript that the appearance of a distinct (gapless) QSL is {\sl not} 
the generic behavior for generalizations of the Kitaev model to other lattice geometries. 
Instead, we show that generically one will find an intermediate regime of enhanced canted magnetic moments in the ground state of AFM Kitaev models, smoothly 
connected to the high-field polarized regime and 
preceded at higher temperatures by a broad regime of proximate spin liquid physics \cite{Banerjee2016proximate,Banerjee2016neutron,Revelli2019fingerprints,Yamaji_clues_2016,Gohlke_dynamics_2017}
i.e. signatures of fractionalization akin to the temperature regime above a true QSL ground state. 
For FM Kitaev models one simply finds a single direct transition from the KSL to the polarized phase.
With regard to the honeycomb Kitaev model, we identify a symmetry mechanism that prevents the formation of a canted 
magnetic regime, consistent with the unusual situation of an intermediate gapless QSL phase there. 

We arrive at these conclusions by studying two representative generalizations of the Kitaev model, for the decorated honeycomb (DH) \footnote{The decorated honeycomb lattice is also sometimes referred to as the Fisher lattice or triangle-honeycomb lattice.} and square-octagon (SO) lattices. 
Both lattices give rise to gapped KSLs in the absence of a field, but they differ in that the DH KSL supports gapless edge modes and non-Abelian anyons \cite{yao-kivelson} while the SO KSL has a gapped edge and Abelian anyons \cite{Yang2007mosaic}. To study the field-driven physics of these Kitaev systems we employ a range of techniques -- exact diagonalization (ED), infinite density matrix renormalization group (iDMRG) calculations and
linear spin-wave theory (LSWT) -- to provide a comprehensive picture of the KSL in the presence of a field. 

The remainder of the manuscript is structured as follows. In Secs.~\ref{sec:Model} and \ref{sec:Low} we review the Kitaev model and the effects of adding an infinitesimal magnetic field. In Sec.~\ref{sec:High} we use LSWT to elucidate the physics of the polarized phase and how the spin wave spectrum behaves as the field is lowered. Sec.~\ref{sec:Int} covers the intermediate-field regime, with the numerical results of our ED and iDMRG computations presented and analyzed. Finally, Sec.~\ref{sec:Disc} provides a comprehensive discussion of all of the results obtained and their place within the wider context of the physics of the Kitaev model and QSLs.


\section{The Kitaev Model}
\label{sec:Model}

\begin{figure} 
\includegraphics[width=\columnwidth]{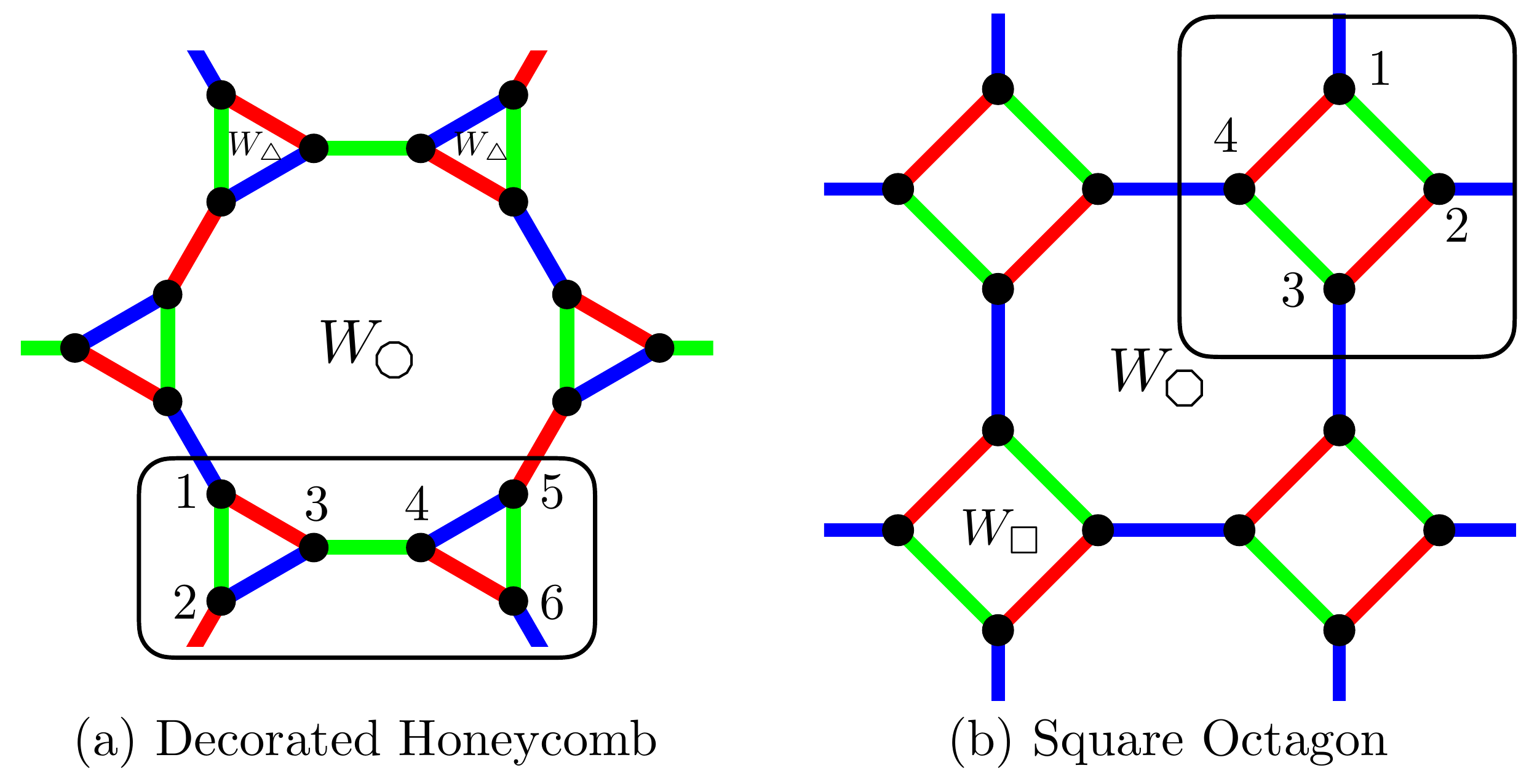}
\caption{\textbf{Lattice structure} of (a) the decorated honeycomb (DH) and (b) square octagon (SO) lattices with unit cells of $6$ and $4$ spins respectively. Each lattice contains two distinct plaquettes, a $3$-site triangular plaquette, $W_\triangle$, and $12$-site dodecagonal plaquette, $W_{\protect\dodecagon{0.5}}$, for the DH lattice and a $4$-site square plaquette, $W_\square$, and $8$-site octagonal plaquette, $W_{\protect\Boctagon{0.5}}$ for the SO lattice.}
\label{fig:Lattices}
\end{figure}

The Kitaev model, originally defined on the honeycomb lattice, can be straightforwardly extended to any tricoordinated lattice with the Hamiltonian
\begin{equation}
H_\text{KSL}^\pm = \pm K \sum_{\avg{i,j}\in\gamma} S_i^\gamma S_j^\gamma
\end{equation}
where $H_\text{KSL}^\pm$ indicates an AFM/FM Kitaev coupling and the three bond directions are denoted by $\gamma\in \left\{ x,y,z \right\}$. Crucially this Hamiltonian has an extensive number of locally conserved plaquette variables $W_p$, $[H_\text{KSL}^\pm , W_p] =0$ and $[W_p,W_{p^\prime}]=0$, defined as the product of bond operators $\prod_{\avg{i,j}\in\gamma} S_i^\gamma S_j^\gamma$ over the bonds of the plaquette. They thus take values $W_p = \pm1$ ($\pm i$) for plaquettes with an even (odd) number of bonds. On each lattice there are $2^{N/2}$ such plaquettes, splitting the Hilbert space into sectors with fixed values of $W_p$ and with $2^N / 2^{N/2} = 2^{N/2} =\left( \sqrt{2}\right)^N$ degrees of freedom in each sector. Note that both the SO and DH lattice have two different types of plaquettes, shown in Fig.~\ref{fig:Lattices}. 

Kitaev showed \cite{Kitaev2006} that by representing the spin operators in terms of Majorana fermion operators it's possible to rewrite the Hamiltonian in terms of a single, free Majorana fermion coupled to a static $\mathbb{Z}_2$ gauge field, whose flux is precisely $W_p$. The ground state flux sector can be determined either numerically, or, in certain cases, via application of Lieb's theorem \cite{Lieb1994flux}. Flux excitations, or ``visons'', occur when one of the $W_p$ are flipped relative to their ground state value. These gauge excitations cost a finite energy, i.e.~the visons are gapped excitations with a gap $\Delta_v \sim \mathcal{O}(K/10)$. On the other hand, the band structure of the free Majorana fermion depends on the lattice geometry. For the original honeycomb case, it consists of gapless Dirac cones \cite{Kitaev2006} while for the DH and SO cases the band structure is gapped \cite{yao-kivelson,Yang2007mosaic}. The KSLs on the SO and DH lattices are thus fully gapped, in both the gauge and matter sectors. 

The intrinsic topological properties of the fully gapped KSLs can be classified according to the Chern number $C$ of their Majorana band structure. The Majorana band structure on the SO lattice is trivial \cite{Yang2007mosaic}, $C=0$, and thus the KSL has the same Abelian topological order as the toric code \cite{Kitaev2003}. On the other hand, on the DH lattice, the Majoranas inhabit a non-trivial band structure \cite{yao-kivelson} with $C=\pm 1$, resulting in non-Abelian Ising anyon topological order. In terms of edge physics, this means that the SO KSL has a trivial gapped edge whereas the DH KSL has a gapless, chiral Majorana edge mode (reflected in a chiral conformal central charge $c=1/2$).  

In the presence of a uniform magnetic field of arbitrary orientation, the Hamiltonian now becomes
\begin{align}
\notag H^\pm &= \pm K \sum_{\avg{i,j}\in\gamma} S_i^\gamma S_j^\gamma - \sum_i \vec{h} \cdot \vec{S}_i \,,\\
H^\pm &= H_\text{KSL}^\pm + H_\vec{h}.
\end{align}
Here, in order to simplify the discussion, we will focus on a field direction that affects all spin components equally, i.e.~a field along the [111] direction, $\vec{h} = (h,h,h)/\sqrt{3}$. For the honeycomb and DH models this direction naturally corresponds to the out-of-plane $c$-axis direction, its relationship to the SO model is discussed in Appendix~\ref{app:canting}.


\section{Methods}
\label{sec:Methods}

To start out, we give a brief overview of the tools used to study the field-induced behavior of the Kitaev model, namely linear spin-wave theory, exact diagonalization and the density matrix renormalization group. 

\subsection{Linear Spin-Wave Theory}

Given a classical ordered state -- such as the polarized state in the high-field limit -- it's possible to look at quantum fluctuations about such a state in a $1/S$ expansion, with $S$ the spin length and $S=\infty$ corresponding to the classical limit. This can be achieved by a standard Holstein-Primakoff transformation, in which the spin operators, at leading order, are replaced by boson creation and annihilation operators:
\begin{align}
S_i^+ \approx \sqrt{2S} b_i \,\, , \,\, S_i^- \approx \sqrt{2S} b_i^\dagger \,\, , \,\, S_i^z \approx S - b_i^\dagger b_i \, ,
\end{align}
for a spin $\vec{S}_i$ aligned along the $z$-axis. For a general classical ordered ground state, one can locally rotate each spin to point along the $z$-axis and then use the above expressions to rewrite the Hamiltonian in terms of the classical ground state energy, at order $S^2$, plus quantum corrections quadratic in the boson operators, at order $S$ (we neglect higher-order corrections). The resulting spin-wave Hamiltonian can be straightforwardly diagonalized, allowing the calculation of the spin-wave spectrum and dynamical spin structure factor using standard procedures \cite{HolsteinPrimakoff,Toth2015}.

\subsection{Exact Diagonalization}

To study the intermediate field regime we use a combination of ED and iDMRG simulations. The ED was numerically implemented using the ARPACK library \cite{arpackusers} and translational symmetry was used in all cases to block diagonalize the Hamiltonian. We studied $N=18, 24$ site clusters for the DH lattice (corresponding to $N_{\text{uc}} = 3,4$ unit cells) and $N=16,24$ site clusters for the SO lattice ($N_{\text{uc}} = 4,6$ unit cells). In both cases the results are qualitatively consistent for the two different system sizes and, as a result, we show throughout the data only for the larger $N=24$ site system. The next largest system sizes, $N_{\text{uc}} = 5$ ($7$) for the DH (SO) lattice, contain a prime number of unit cells and thus can only be realized in an $L_x \times L_y$ geometry as a highly anisotropic, essentially 1-d system, $5 \times 1$ ($7\times 1$). Such geometries are not suitable for our study and, beyond that, larger system sizes are beyond the scope of the present study. The finite temperature ED results were calculated using the method of thermal pure quantum states \cite{TPQS2012,CanTPQS2013}. 

\subsection{Density Matrix Renormalization Group}

We complement ED by employing iDMRG %
\cite{white_density_1992,mcculloch_infinite_2008,phien_infinite_2012}
and representing the quantum many body wave function, i.e. the ground state, 
as a matrix product states (MPS).
While initially developed for one-dimensional systems, MPS and iDMRG have been proven to be fairly unbiased and well controlled even for two-dimensional systems, which can be accessed by wrapping its lattice on a cylinder and winding a chain along the sites.
 Translational symmetry allows to study cylinders whose length is in the thermodynamic limit, while the circumference remains finite.
This results in lines of accessible momenta in reciprocal space. 
Here, we use various geometries (DH, DH-2, SO, and SO-2) with $L_y = 3$ and $4$ unit cells along the circumference.
The geometries and their accessible momenta are illustrated in Fig.~\ref{fig:dmrg_geometries} in Appendix \ref{app:dmrg_geometries}. 
We use a bond dimension of up to $\chi = 1200$ for the ground state MPS.

A matrix product operator based time-evolution (tMPO) \cite{zaletel_timeevolving_2015} enables us to compute the dynamical spin structure factor as the spatio-temporal Fourier transform of the dynamical correlation function
$C^{\gamma\gamma}_{ij} = \langle \Psi_0| S^\gamma_i U(t) S^\gamma_j |\Psi_0\rangle$, where $|\Psi_0\rangle$ is the ground state.
The time-evolution operator for discrete time-steps $U(\Delta t)$ is hereby represented as a matrix product operator and iteratively applied to the MPS.
Due its computational cost, we compute the time-evolution only using \emph{SO} and \emph{DH} geometry with $L_y=3$, and limit the bond-dimension to $\chi=256$.


\section{Low-Field Regime: $h \ll K$}
\label{sec:Low}

First we review the physics of the low-field regime, $h\rightarrow 0$, which is already well-known in the literature for the two lattice geometries at hand \cite{yao-kivelson,Yang2007mosaic} and, due to the exact solvability of the Kitaev model, has been explored in a variety of other lattice geometries in two \cite{Kells2011Kaleidoscope} and three \cite{Si2008anyonic,Mandal2009exactly,Hermanns2014quantum,Kimchi2014three,Hermanns2015weyl,Hermanns2015spin-peierls,Obrien2016classification,Yamada2017,Eschmann2020classification} spatial dimensions. Here, concentrating on the two principal lattice geometries at hand, we will remark first on the impact on the flux degrees of freedom, and then the Majorana sector. 

\subsection{Flux Degrees of Freedom}

In the presence of a magnetic field the plaquette variables $W_p$, which correspond to the flux of the $\mathbb{Z}_2$ gauge field, are no longer conserved $[H_{KSL}^\pm, W_p] \neq 0$. This means that the gauge field can no longer be considered static, but instead becomes dynamic, with the visons now mobile excitations with a non-zero bandwidth. The exact solvability of the model is lost, with the Hilbert space no longer separable into well-defined flux sectors. 

Though the model is no longer exactly solvable, we can still make a definite statement on the stability of the KSL in the presence of the field. The flux gap, which is finite in the absence of a field, ensures that the KSL is stable to infinitesimal perturbations, including a magnetic field. There will thus be a \emph{finite} extent of the KSL in the field-dependent phase diagram and a finite $h_c^{\text{KSL}} > 0$ at which it is destroyed. 

\subsection{Majorana Degrees of Freedom}

To investigate the impact on the Majorana degrees of freedom, Kitaev performed perturbation theory about the exactly solvable point with $H_0 = H_\text{KSL}^\pm$ and the magnetic field term as a perturbation $V = H_\vec{h}$ \cite{Kitaev2006}. At linear order in $h$, there is no contribution. At second order there is a non-zero contribution that simply renormalizes the value of the Kitaev interaction (and crucially does not break TRS). At third order, there are two distinct types of contributions, one of which can be rewritten in terms of Majorana fermions as a quadratic next-nearest neighbor (NNN) hopping term (the other cannot be simplified, breaking the integrability of the model). Adding just this NNN hopping term, which is proportional to $(h_x h_y h_z)$, the model remains exactly solvable and explicitly breaks TRS. In the language of the spin degrees of freedom adding this term leads to
\begin{align}
H^\pm &= \pm K \sum_{\avg{i,j}\in\gamma} S_i^\gamma S_j^\gamma -  \kappa \sum_{ijk} S^x_i S^y_j S^z_k  \, ,
\end{align}
where $ijk$ are NNN sites and $\kappa \sim (h_x h_y h_z) / ( \Delta_v^{ij} \Delta_v^{jk} )$, with $\Delta_v^{ij}$ the flux gap due to creating visons on the two plaquettes neighboring the $ij$-bond. 

Note that, for the perturbation theory to be valid, it's necessary not for the field to be small compared to the Kitaev coupling $K$, but small compared to the flux gap $\Delta_v$. Since $\Delta_v \sim \mathcal{O}( K/10)$ for the zero-field KSL this means that there is a limited range of fields in which perturbation theory should be trusted, $h \ll \mathcal{O}( K/10)$.  

For the honeycomb case, the addition of the NNN hopping term for the Majorana fermions gaps out the Dirac cones, creating a fully gapped QSL with non-Abelian, Ising anyon topological order \cite{Kitaev2006}. However, for the DH and SO cases, as the band structure is already gapped at zero-field, the additional term does not qualitatively alter the physics: The DH also exhibits an extended phase with Ising topological order, while the SO shows a stable phase with Abelian topological order. This is schematically summarized in Fig.~\ref{fig:LowField}.
The impact of a small magnetic field, within the context of the perturbation theory outlined above, on KSLs for other tricoodinated lattices has been discussed in Refs.~\onlinecite{Hermanns2015spin-peierls,Obrien2016classification}. 

\subsection{Summary}

\begin{figure}  
\includegraphics[width=\columnwidth]{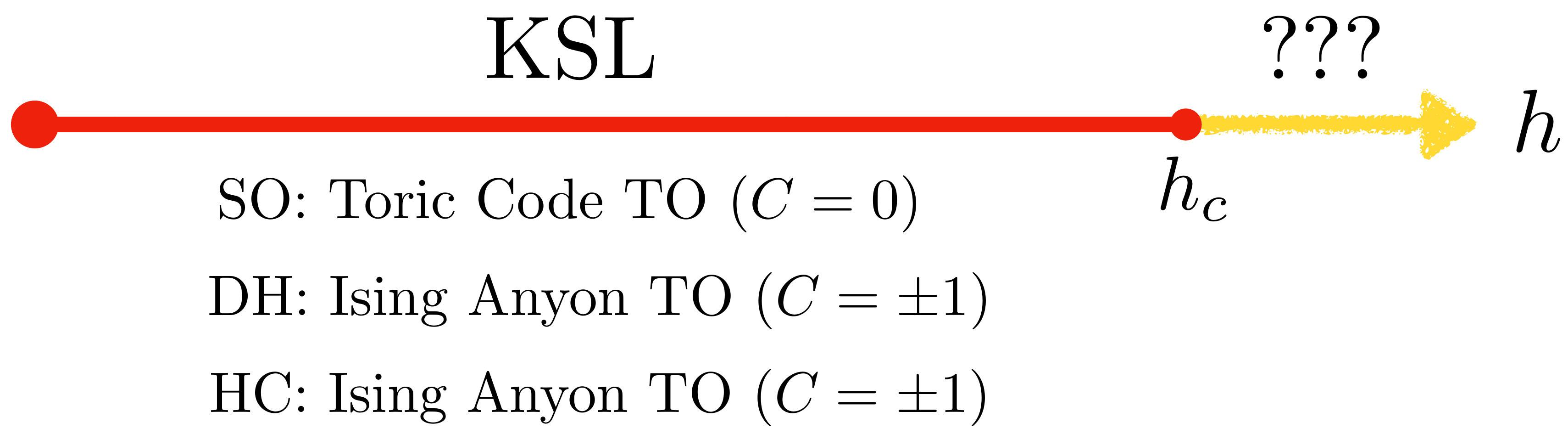}
\caption{\textbf{Generic low-field behavior} of the KSL for the FM/AFM cases. It is stable up to a finite critical field $h_c$, with the topological order for the square-octagon (SO), decorated honeycomb (DH) and honeycomb (HC) lattices listed as examples of low-field KSLs (the Chern number $C$ of the Majorana bands are also given).}
\label{fig:LowField}
\end{figure}

In the low-field regime we have a finite extent of the KSL, due to the finite flux gap $\Delta_v$. Hence, all lattices have a finite critical field $h_c^\text{KSL}$ at which the KSL is destroyed. Using perturbation theory, for $h\ll \Delta_v$, one can qualitatively investigate the effects of the field on the Majorana band structure by adding one of the terms that appears that breaks TRS but retains the exact solvability of the model, a NNN hopping term in the Majorana description. For the SO and DH lattices there is no qualitative change in the band structure, and thus the KSL. The situation is summarized in Fig.~\ref{fig:LowField}.

\section{High-Field Regime: $h \gg K$}
\label{sec:High}

In the high-field regime the system will always form a partially polarized state, independent of the underlying lattice. The lack of any spin rotational symmetry in the Kitaev model means that the ideal fully polarized state, $\ket{\uparrow\uparrow\dots}$, with a fully saturated magnetization, is not an eigenstate of the Hamiltonian. It is only as $h\rightarrow\infty$ that the system approaches this ideal product state. Quantum fluctuations are suppressed with increasing field and we can do a spin-wave expansion around the classical polarized state.  

We will restrict ourselves to linear spin-wave theory, using the standard Holstein-Primakoff expansion about the classical fully polarized state. Such a study has already been done for the Kitaev model on the honeycomb lattice \cite{McClarty2018topological}, which found the appearance of topological magnons (with their bands carrying a non-trivial Chern number), as well as extended Kitaev models on the honeycomb \cite{Janssen2016,Janssen2017,consoli2020} and hyperhoneycomb lattices \cite{SungBin2014,Kruger2020}. Here, we investigate what happens for the pure Kitaev model on other tricoordinated lattices, with a focus on the DH and SO lattices. In particular, we are interested in what happens as the field is decreased, when does LSWT fail? How do the bands evolve with field? 

\begin{figure}[b]  
\includegraphics[width=\columnwidth]{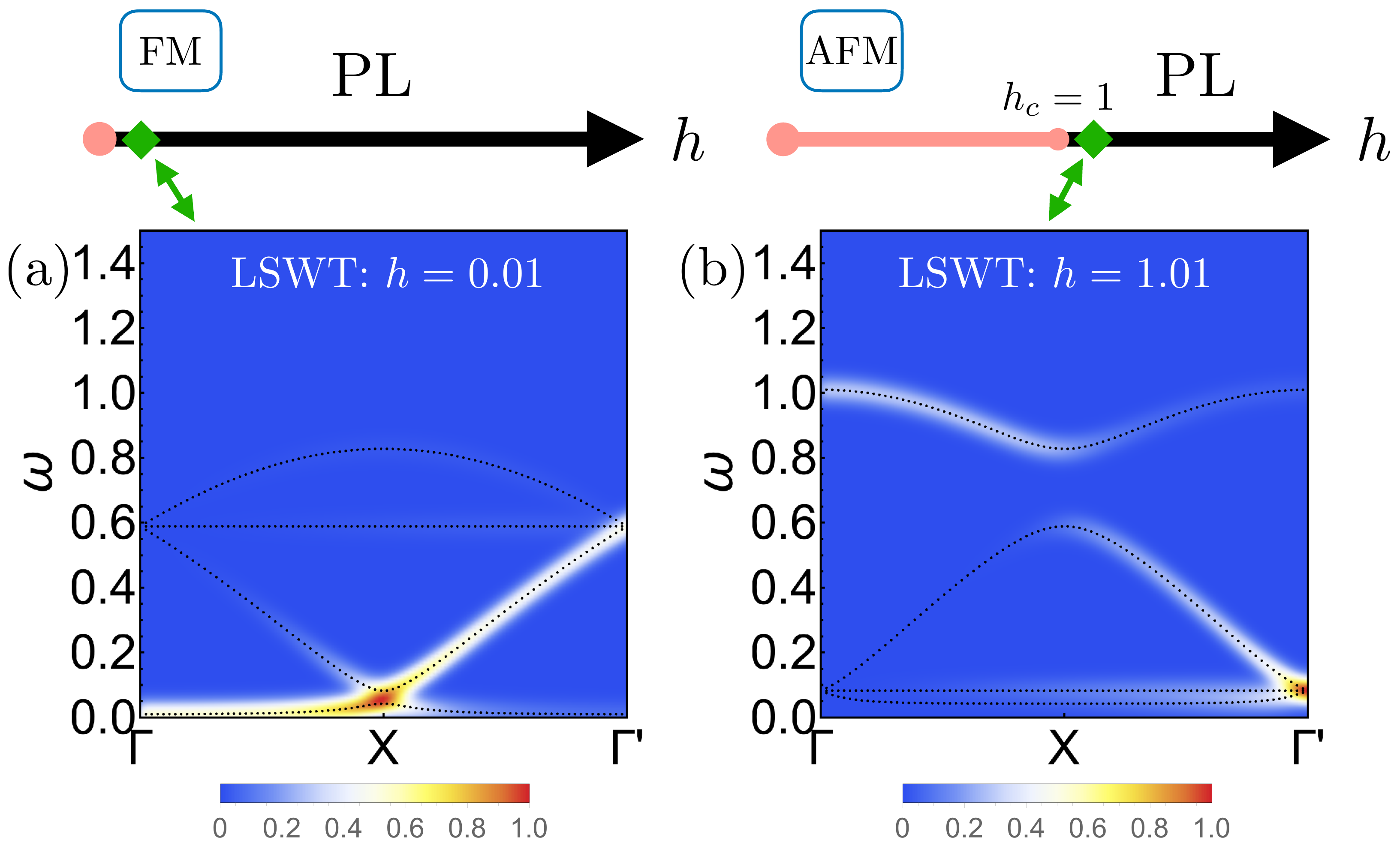}
\caption{\textbf{Generic high-field behavior} of the FM/AFM Kitaev models.
		Shown is the dynamical spin structure factor for the SO model,
		with panel (a) showing the FM case at $h=0.01$ and
		panel (b) showing the AFM case at $h=1.01$.
		The dotted black lines indicate the spin wave bands (intensity is in arbitrary units).} 
\label{fig:HighField}
\end{figure}

\subsection{Results: FM Model}

The first step in LSWT is to determine the classical ground states of the model about which we do the spin-wave expansion. For the FM Kitaev model the classical fully polarized state, with all spins pointing along the direction of the applied field, is the classical ground state all the way down to zero field. That this is true can be easily seen starting from the zero-field limit. In this limit the ground state is a classical spin liquid with a large manifold of degenerate states \cite{Chandra2010classical} (giving rise to a Coulomb phase \cite{Sela2014order,Henley2010coulomb}). Included within this manifold is the fully polarized state. As a result, applying a field immediately lowers the energy of the polarized state relative to the other classical states within the manifold and it becomes the classical ground state. It is thus possible to carry out LSWT all the way down to zero field. 

For the Kitaev model, the same qualitative behavior can be seen in all 2d/3d lattices, as summarized in Fig~\ref{fig:HighField}. The bandwidth stays constant as a function of field and the spin gap goes to zero as the field goes to zero. There is a flat band(s) that hits zero energy at zero field. 
As an example, the dynamical spin structure factor at $h=0.01$ is shown in Fig.~\ref{fig:HighField}(a) for the SO lattice where one can clearly identify an almost flat band just above zero energy.  

We know, from the previous section, that the KSL is stable to a finite critical field so the LSWT prediction of a stable polarized state all the way to $h=0$ is clearly incorrect. Nevertheless, the fact that LSWT can be applied down to zero field suggests that, generically, for the FM Kitaev model, there will only be a small field window at low fields within which interesting quantum effects will occur. 

\subsection{Results: AFM Model}

\begin{figure}  
\includegraphics[width=\columnwidth]{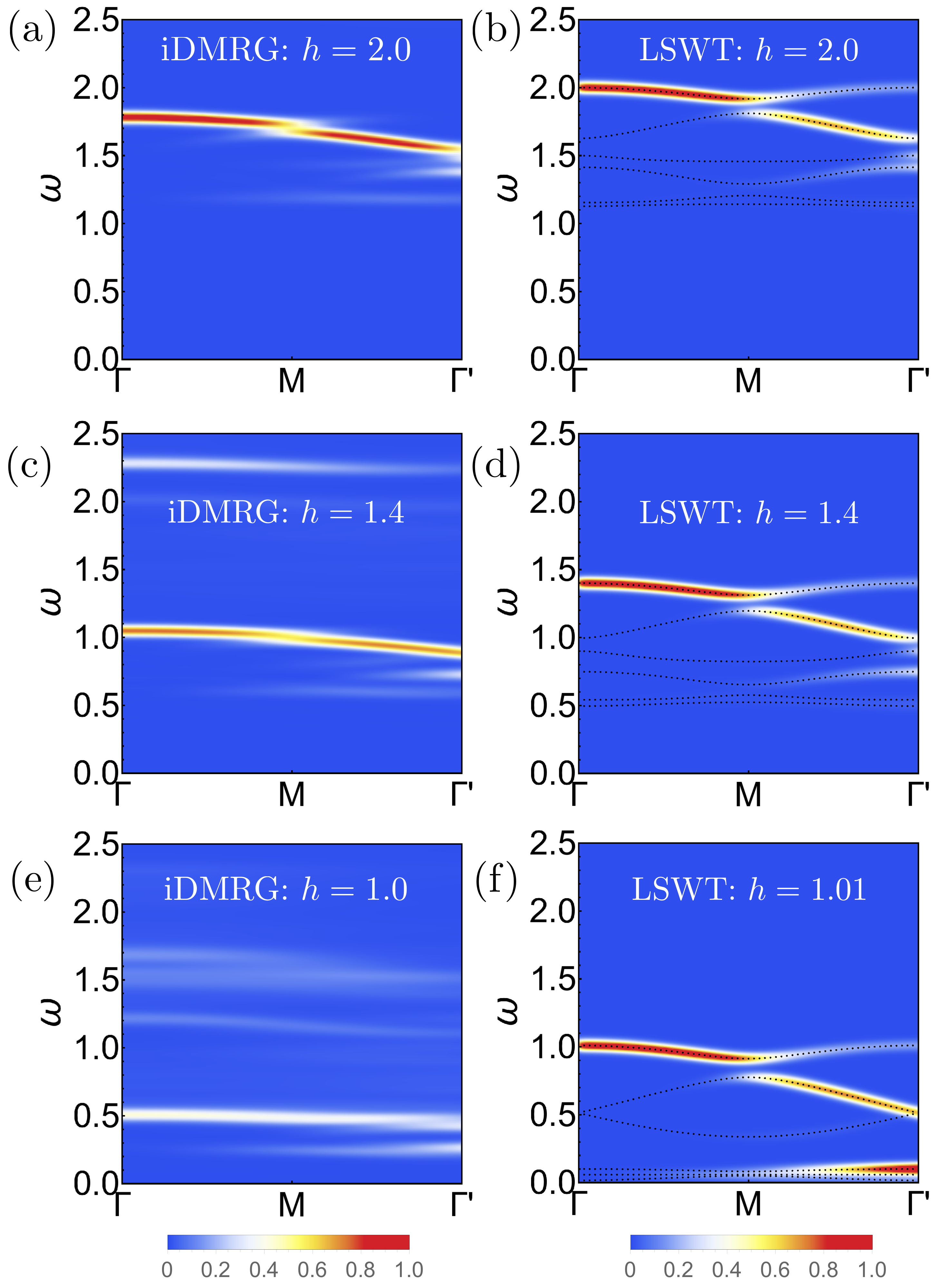}
\caption{\textbf{Comparison of the dynamical spin structure factor} for the DH lattice within the polarized phase, as computed by iDMRG in (a), (c) and (e), and LSWT in (b), (d) and (f). The dotted black lines in indicate the spin wave bands (the intensity is in arbitrary units).}
\label{fig:YK_DSF_Comp}
\end{figure}

For the AFM model the zero-field classical ground state manifold does not include the polarized state (such a state has FM spin-spin correlations and so clearly would not be a ground state of the AFM model). As a result an infinitesimal field is not expected to polarize the system. Indeed, the fully polarized state is only the classical ground state for $h>2S$. Below this, $h<2S$, there is no simple ordered classical ground state but instead a complex ground-state manifold which cannot be treated within simple LSWT (see Appendix \ref{app:classical_kitaev} for further details). For the spin-1/2 problem at hand LSWT about the polarized state is thus restricted to $h>1$.  

Again we see the same qualitative behavior in all 2d/3d lattices. Similar to the FM case the bandwidth is constant and the spin gap goes to zero via a flat band(s) hitting zero energy, except this time at $h_c=1$, rather than $h=0$ as in the FM case. Fig.~\ref{fig:HighField}(b) shows the dynamical spin structure factor for the AFM SO model at $h=1.01$, where again one can clearly observe, in this case, two almost flat bands just above zero energy.  

We note that non-linear corrections beyond LSWT will change the critical field value $h_c=1$ at which the polarized phase becomes unstable, indeed this has been shown explicitly in the honeycomb case \cite{McClarty2018topological,consoli2020}. This failure of LSWT can also be clearly seen in Fig.~\ref{fig:YK_DSF_Comp}, in which we compare the dynamical structure factor for the DH lattice calculated using iDMRG and LSWT. At high fields, we see a nice agreement between the two methods, as shown in Fig.~\ref{fig:YK_DSF_Comp}(a) and (b) for $h=2.0$, due to the suppression of quantum fluctuations and mixing with the two-particle continuum. However, as the field is lowered, the methods begin to diverge, at $h=1.4$ in (c) and (d) the total bandwidth is noticeably smaller in iDMRG and weight begins to appear at higher energies. At $h=1.0, 1.01$ in (c), (d) the lowest three bands in LSWT are almost at zero energy and flat across the BZ, signalling the impending instability of the polarized state, whereas in iDMRG there is still a sizeable spin gap $\sim 0.2 K$ and also a significant region of continuum excitations. The polarized phase thus exists over a wider field range than LSWT would suggest.   

The fact that LSWT about the polarized state fails at a finite critical field is at least, unlike the FM case, consistent with the finite extent of the KSL phase. Indeed, compared to the FM case, this failure leaves a substantial field range within which new quantum phenomena may occur. 

\subsection{Summary}

There is a universal behavior to the classical fully polarized state of the 2d/3d Kitaev models as field is reduced within LSWT. The magnon bands move to lower and lower energies and a single, or multiple, flat bands hit zero energy at a critical field $h_c=0$ for the FM case and $h_c = 1$ for the AFM case (for $S=1/2$). Going beyond LSWT to the fully quantum model we thus expect that there is a much wider region of parameter space for interesting physics to emerge in the AFM case, as compared to the FM case.


\section{Intermediate-Field Regime: $h\sim K$}
\label{sec:Int}

So far we have seen that the KSL must be stable up to some finite critical field $h_c$ and that, within LSWT, the high-field polarized phase becomes unstable at $h=0$ ($h=1$) for the FM (AFM) model. Now we move on to examine the full phase diagram for the specific case of the DH and SO lattices, and in particular explore what happens at intermediate fields, away from the simple limits studied in the previous two sections. For this, we turn to unbiased numerical techniques, namely ED and iDMRG, to go beyond the perturbative and semi-classical regimes of the previous sections.   

\subsection{Results: FM Model}

First we take a look at the FM Kitaev models, which turn out to show rather straightforward behavior in the presence of an applied field. 

\subsubsection{Decorated-Honeycomb Lattice}

\begin{figure}[h!]
\includegraphics[width=\columnwidth]{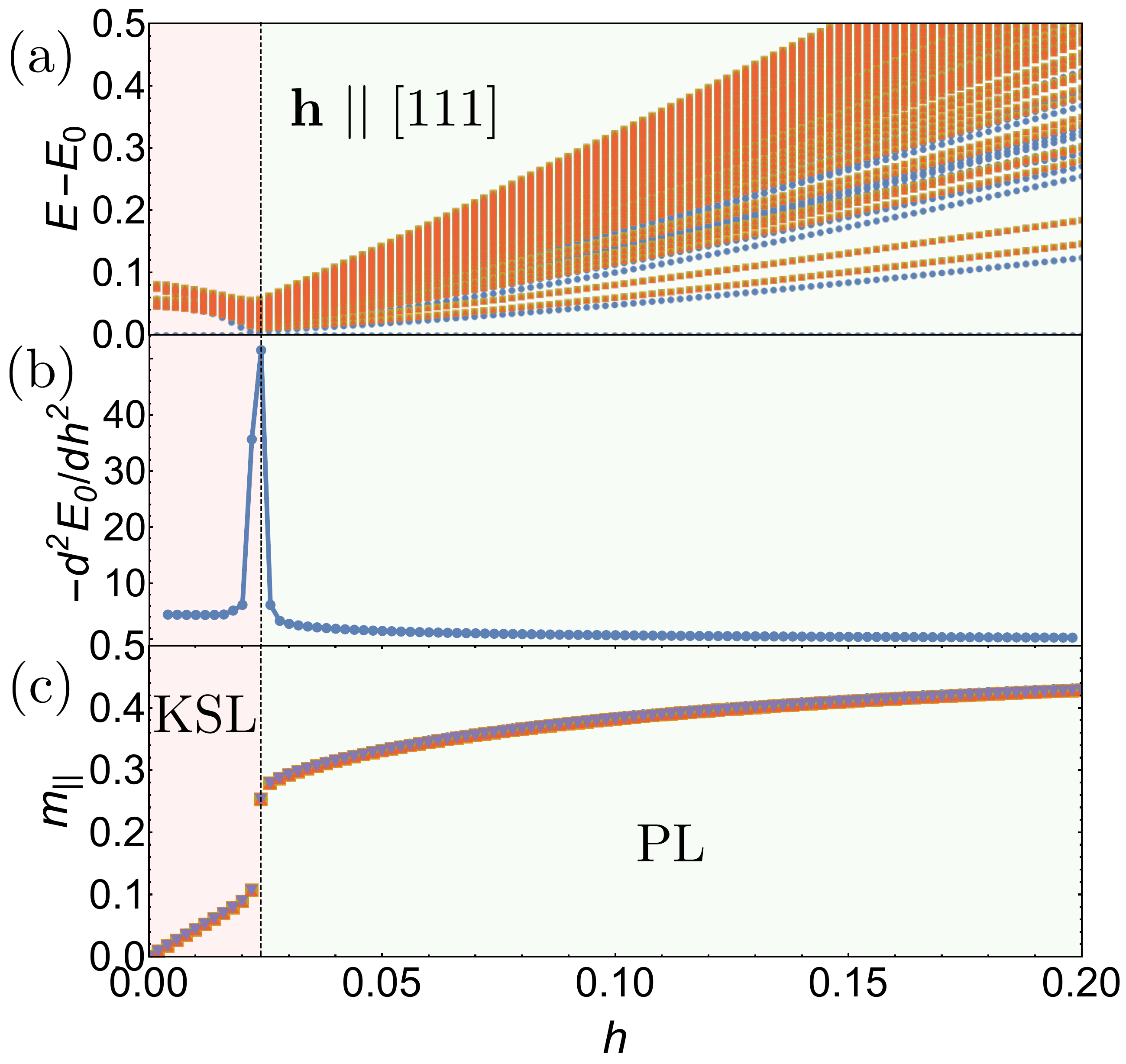}
\caption{\textbf{ED results for the FM Kitaev model on the DH lattice} with (a) energy spectrum (different colored points correspond to different momentum sectors), (b) second derivative of the ground state energy, $d^2E_0/dh^2$, and (c) magnetization parallel to the [111] field direction, $m_{||}$ (the 6 different markers correspond to the 6 sites in the unit cell).}
\label{fig:FM_YK_ED_PD}
\end{figure}

The ED energy spectrum for the DH lattice for an $N=24$ site cluster is shown in Fig.~\ref{fig:FM_YK_ED_PD}(a). There is a clear transition at $h_c^{\text{KSL}}=0.024$, which occurs via a level crossing, between the low-field KSL and the high-field polarized state. The transition is reflected in a sharp peak in the second derivative of the ground state energy, shown in Fig.~\ref{fig:FM_YK_ED_PD}(b) (as well as a sharp drop in the ground state fidelity, shown in Appendix \ref{app:gs_fidelity}). Above the critical field the lowest-lying energy levels exhibit a linear increase in energy with increasing field, as expected for the lowest-lying magnon excitations of the polarized state. The magnetization parallel to the field, $m_{||}$, shown in Fig.~\ref{fig:FM_YK_ED_PD}(c), jumps at the transition and then smoothly increases as field increases, reaching, for example, $\sim 86\%$ of its saturated value of $1/2$ at a field of $h=0.2$. 

\begin{figure}[h!]
\includegraphics[width=\columnwidth]{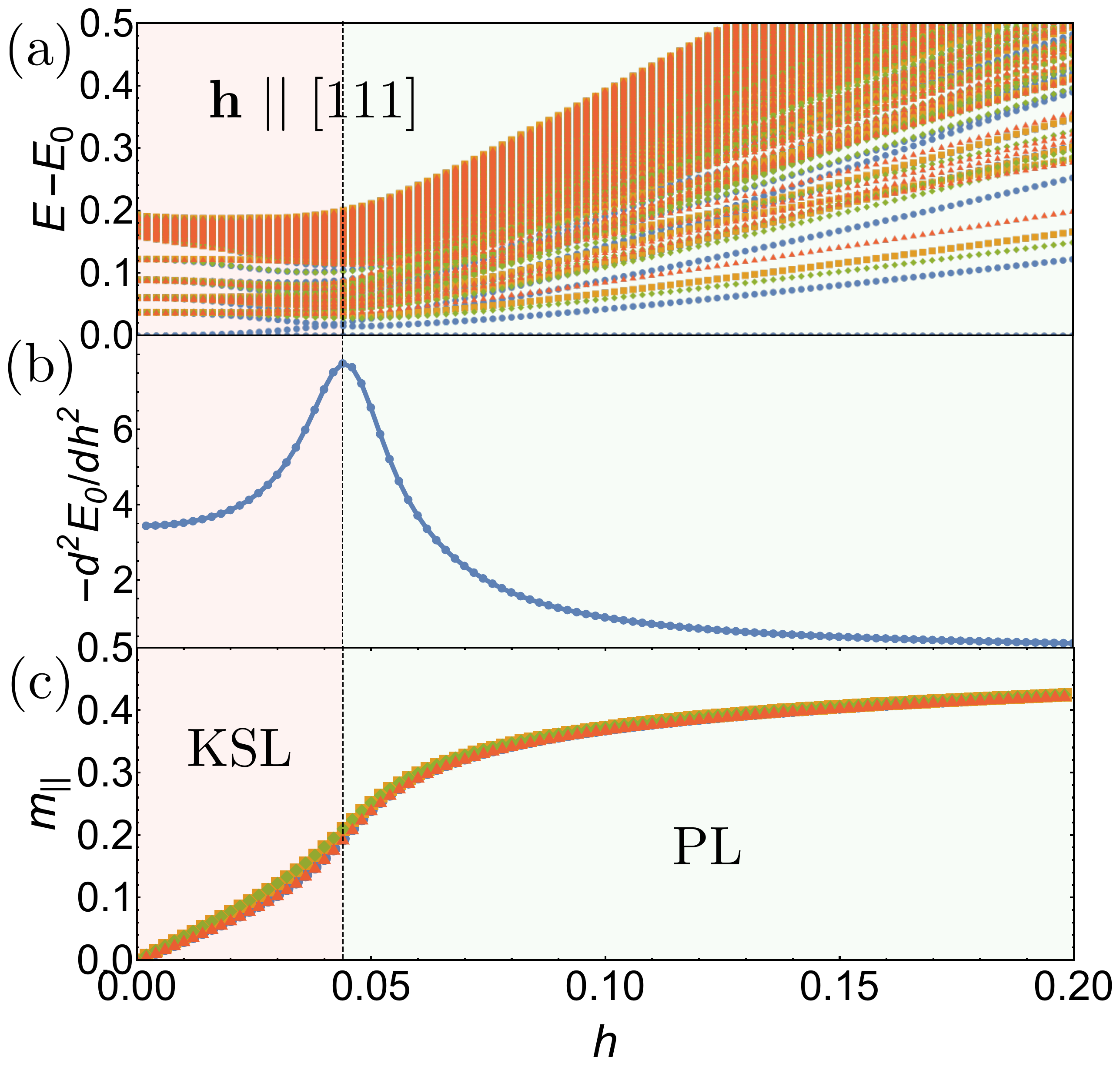}
\caption{\textbf{ED results for the FM Kitaev model on the SO lattice} with (a) energy spectrum (different colored points correspond to different momentum sectors), (b) second derivative of the ground state energy, $d^2E_0/dh^2$, and (c) magnetization parallel to the [111] field direction, $m_{||}$ (the 4 different markers correspond to the 4 sites in the unit cell).}
\label{fig:FM_SO_ED_PD}
\end{figure}

\subsubsection{Square-Octagon Lattice}

The phase diagram for the SO lattice is qualitatively similar to the DH case. The energy spectrum and ground state energy second derivative, shown in Fig.~\ref{fig:FM_SO_ED_PD}(a) and (b), again clearly point to a single transition from the low-field KSL to the high-field polarized phase, in this case at a field of $h_c^{\text{KSL}}=0.044$ (with, again, the transition also reflected in the ground state fidelity shown in Appendix \ref{app:gs_fidelity}). The low-lying energy levels show a clear linear increase in field above $h_c^{\text{KSL}}$, with the plaquette flux on the squares/octagons smoothly going to zero and the magnetisation parallel to the field, $m_{||}$ shown in Fig.~\ref{fig:FM_SO_ED_PD}(c), smoothly approaching saturation as the field is increased further, e.g. $W_p=-0.09$ ($-0.05$) for squares (octagons) and $m_{||}=0.42$ at $h=0.2$.  

\subsection{Results: AFM Model}

Turning now to the AFM models, we discuss their phase diagrams and the non-trivial behavior that arises within the intermediate field regime.  

\subsubsection{Decorated-Honeycomb Lattice}

\begin{figure}
\includegraphics[width=\columnwidth]{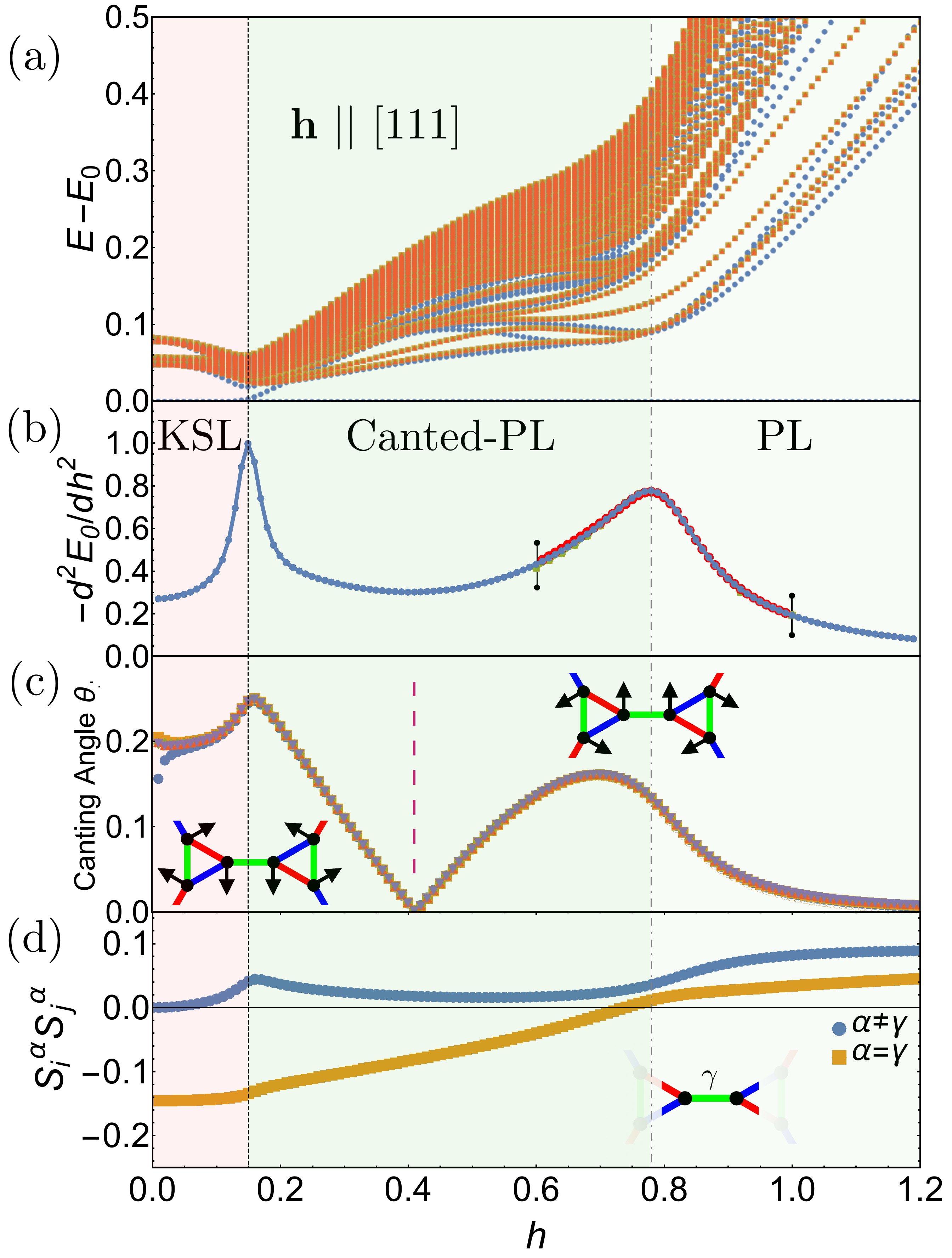}
\caption{\textbf{ED results for the AFM Kitaev model on the DH lattice} with (a) energy spectrum (different colored points correspond to different momentum sectors), (b) second derivative of the ground state energy, $d^2E_0/dh^2$ (note that in the marked region $h=0.6-1.0$ additional ED and iDMRG data is shown), (c) canting angle $\theta_\perp$ of the local moments $\avg{\vec{S}_i}$ away from the field direction (the 6 different markers correspond to the 6 sites in the unit cell) and (d) NN spin-spin correlations on the bonds connecting the triangular plaquettes.}
\label{fig:YK_ED_PD}
\end{figure}

The ED energy spectrum for the $N=24$ site DH lattice cluster is shown in Fig.~\ref{fig:YK_ED_PD}(a). There is a clear transition out of the KSL at a field of $h_c^{\text{KSL}} = 0.15$. However, unlike the FM case, there is an intermediate regime visible, between $h_c^{\text{KSL}}=0.15$ and $h=0.78$, before the spectrum bends and the lowest-lying states start to exhibit the linear-in-field behavior of the high-field polarized state. This intermediate regime can also be observed in the second derivative of the ground state energy, shown in Fig.~\ref{fig:YK_ED_PD}(b). There, the sharp peak at $h_c^{\text{KSL}}$ is followed by a rather broad peak at higher fields, centered at $h=0.78$. Indeed, in the field range $0.6-1.0$, Fig.~\ref{fig:YK_ED_PD}(b) actually includes $d^2E_0/dh^2$ data from both ED, $N=18, 24$ site clusters, and iDMRG, cylinders of width $L_y=3,4$. All of the curves lie on top of one another, indicating that this higher-field peak is completely independent of system size and geometry. Its broadness in field, with a width in field of $\sim \!\! 0.2$, does not fit the behavior expected for a first-order transition and the lack of any kind of scaling with system size or geometry does not fit the behavior of a continuous second-order transition either. Instead, such behavior suggests a crossover, rather than a true phase transition, from the intermediate-field region to the high-field polarized region.  

\begin{figure}
\includegraphics[width=\columnwidth]{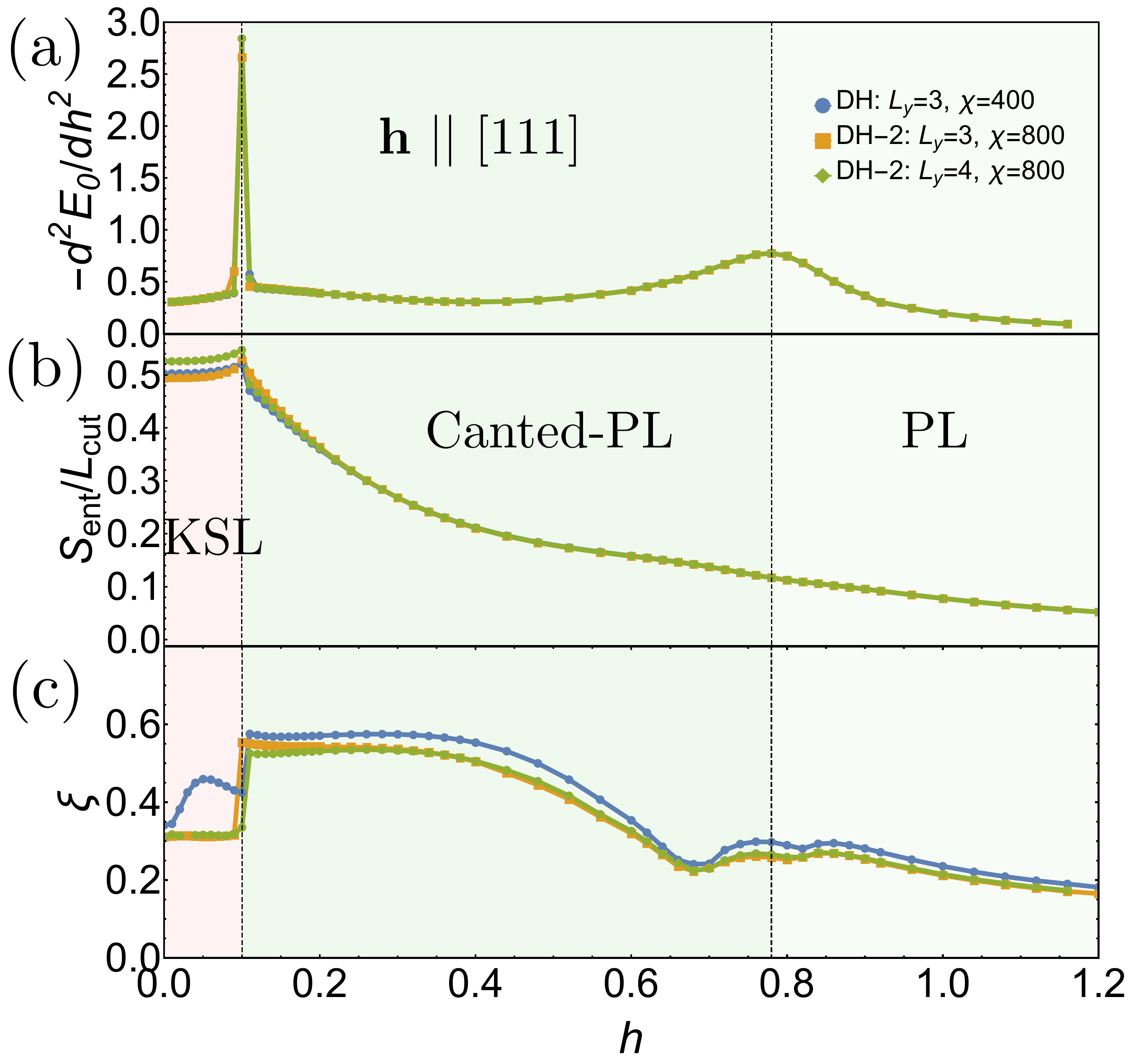}
\caption{\textbf{iDMRG results for the AFM Kitaev model on the DH lattice} showing (a) second derivative of the ground state energy, $d^2E_0/dh^2$, (b) entanglement entropy $S_{\text{ent}}/L_{\text{cut}}$ and (c) the extracted correlation length.}
\label{fig:YK_DMRG_PD}
\end{figure}

What quantitative changes occur when crossing to the intermediate-field region? Fig.~\ref{fig:YK_ED_PD}(c) shows the canting angle $\theta_\perp$, the angle between the local magnetic moments $\avg{\vec{S}_i}$ and the $[111]$ field direction, from ED as a function of applied field. Approaching from high fields the spins begin to cant away from the field direction, with $\theta_\perp$ reaching a local maximum just beyond the crossover and then decreasing and going to zero mid-way within the intermediate-field region. This zero of the canting angle signals a switch from one form of symmetry-allowed canting pattern to another, with the relevant patterns shown in Fig.~\ref{fig:YK_ED_PD}(c) (for more details see Appendix \ref{app:canting}). We thus refer to this intermediate-field region as the ``canted-polarized'' regime. Finally, in Fig.~\ref{fig:YK_ED_PD}(d), the NN spin-spin correlations $\braket{S_i^\gamma S_j^\gamma}$ for $\gamma$-bonds connecting triangles switch from being FM at high fields to AFM at the crossover field. Note that all of these results are fully consistent with iDMRG, though there, because the cylinder geometries used do not support all lattice symmetries of the Hamiltonian, the two canting patterns at low and high-field are different to those two encountered in ED. 

The trivial nature of the canted-PL regime is further reflected in the iDMRG data of Fig.~\ref{fig:YK_DMRG_PD}. The extent of the KSL region is slightly smaller, with $h_c^{\text{KSL}} = 0.10$ in  Fig.~\ref{fig:YK_DMRG_PD}(a), as compared to $h_c^{\text{KSL}} = 0.15$ in the ED data of Fig.~\ref{fig:YK_ED_PD}(b). Within the canted-PL regime, the entanglement entropy, $S_{\text{ent}}$, displays the characteristic area-law behavior expected for a gapped state. This is reflected in the fact that $S_{\text{ent}}/L_{\text{cut}}$, shown in Fig.~\ref{fig:YK_DMRG_PD}(b), does not scale with cylinder width. In other words, $S_{\text{ent}}/L_{\text{cut}}$ is constant and thus $S_{\text{ent}} \propto L_{\text{cut}}$, as expected for a trivial gapped phase. $S_{\text{ent}}$ does not exhibit any signature of the crossover between the canted-PL and high-field PL regimes, nor does the correlation length exhibit any significant increase. Combined with the ED results, we thus get a clear picture of the ground state phase diagram. There is just a single phase transition with the low-field KSL phase giving way to a canted-PL regime, which is then smoothly connected, via a crossover, to the high-field PL phase.     

\begin{figure}[h!]
\includegraphics[width=\columnwidth]{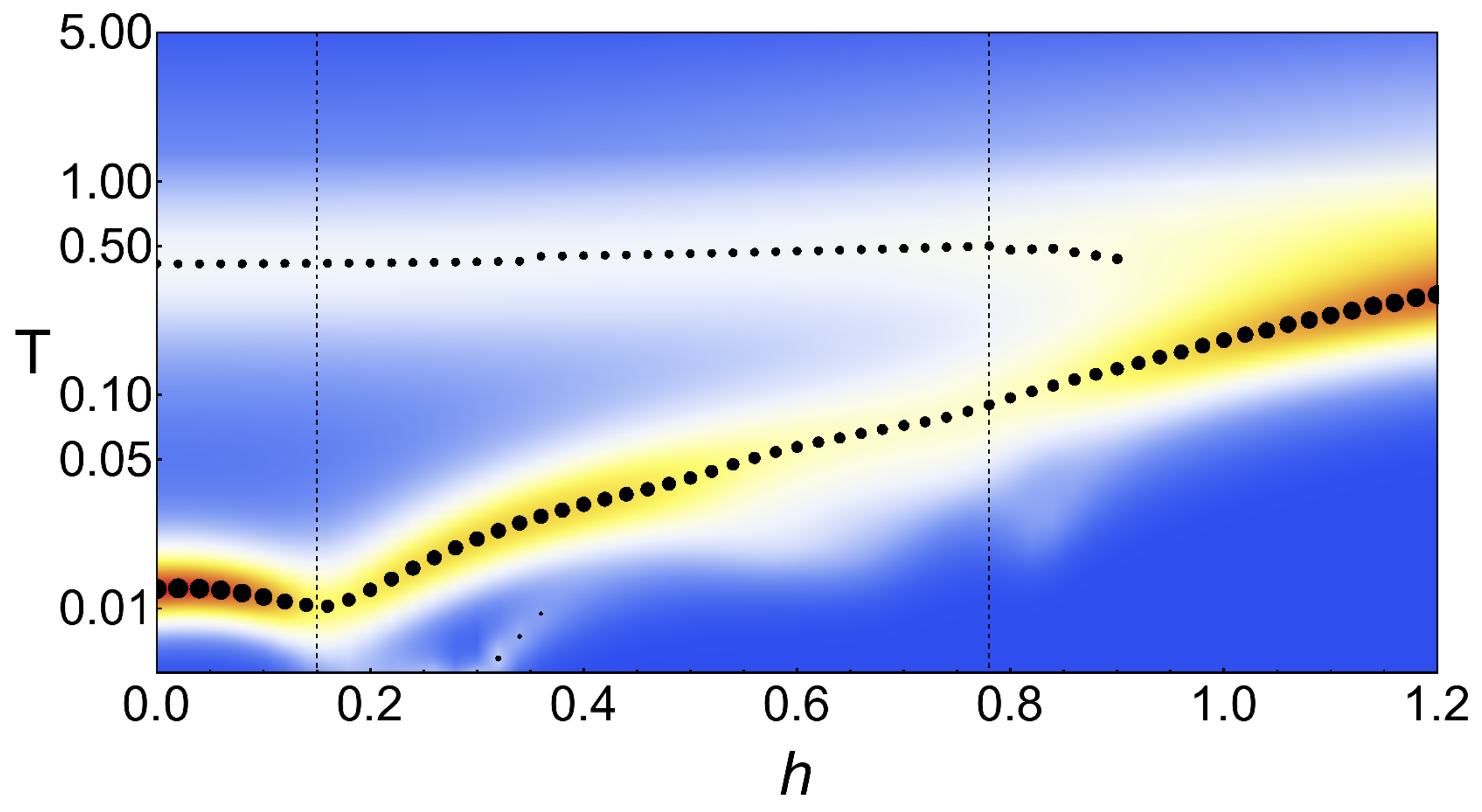}
\caption{\textbf{Specific heat for the DH lattice} as a function of applied field calculated using ED. The black dots denote peaks in the specific heat, with their size proportional to the height of the peak.}
\label{fig:YK_Cv}
\end{figure}

The finite temperature behavior of the canted-PL regime can be seen in the specific heat data, calculated with ED using the thermal pure quantum states method \cite{TPQS2012,CanTPQS2013}, in Fig.~\ref{fig:YK_Cv}. In the KSL there is a signature two-peak structure \cite{Nasu2014vaporization} to the specific heat, associated with the onset of fractionalization at the higher-$T$ peak, and the ordering of the plaquettes at the lower-$T$ peak. Once the transition out of the KSL is crossed the lower peak begins to move to higher and higher temperature, eventually merging with the higher-T peak to form a single peak, as expected for a trivial gapped phase, at a field just above the crossover to the high-field PL regime.  We further comment on the thermodynamics of the intermediate phase, in particular its relation to proximate spin liquid behavior, in the discussion section below. 

\subsubsection{Square-Octagon Lattice}

\begin{figure}[b]
\includegraphics[width=0.875\columnwidth]{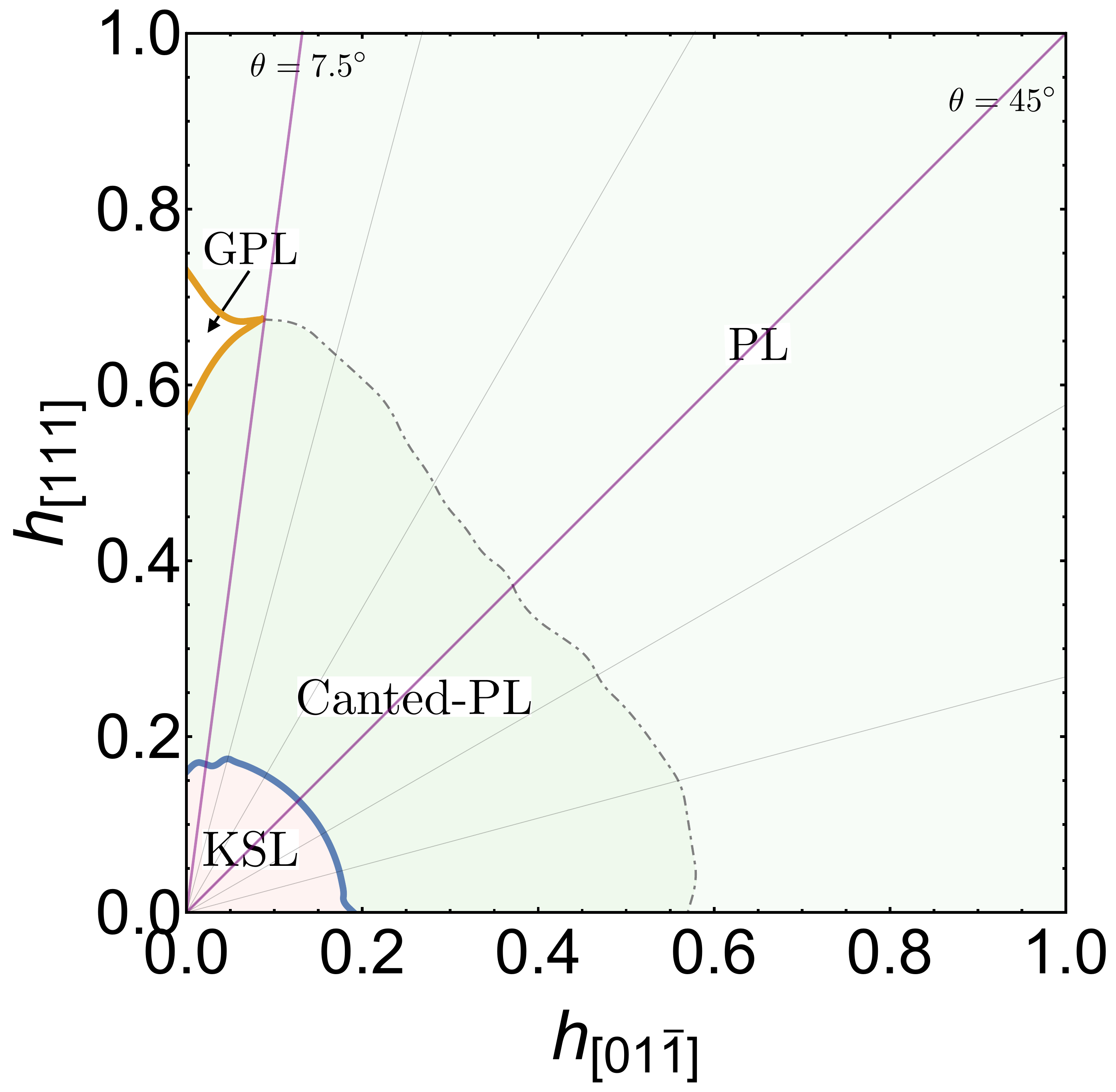}
\caption{\textbf{Tilted field phase diagram for the AFM Kitaev model on the SO lattice} with respect to fields tilted between the $ [111]$ and $[0\bar{1}1]$ directions.}
\label{fig:SO_2dPD}
\end{figure}

To discuss the in-field physics for the SO lattice, it is helpful to first consider more general, tilted field directions between the $[111]$ and, for example, the $[0\bar{1}1]$ 
directions. Much of the key physics readily reveals itself for such tilted field directions, while the $[111]$ field direction turns out to be somewhat special. 
In order to set the stage for this discussion, Fig.~\ref{fig:SO_2dPD} shows the two-dimensional phase diagram from ED for field directions varying from the $[111]$ to the $[0\bar{1}1]$ direction, with the field direction parameterized by a tilting angle $\theta$ via $\vec{h} = h \left(\cos\theta \,\vec{\hat{h}}_{[111]} + \sin\theta \,\vec{\hat{h}}_{[0\bar{1}1]}\right)$. For the vast majority of tilting angles, $\theta \gtrsim 7^\circ$, we find a phase diagram qualitatively similar to the DH case, with the KSL at low fields, followed by a phase transition to an intermediate canted-PL regime, and then a crossover to the high-field PL regime. For small tilt angles, i.e.~fields close to the [111] direction, we observe a new phase that emerges between the canted-PL and PL regimes. First though we focus on a cut at $45^\circ$, to examine the physics that dominates the majority of the phase diagram, before then discussing the special case of small tilt angles. 

\begin{figure}
\includegraphics[width=\columnwidth]{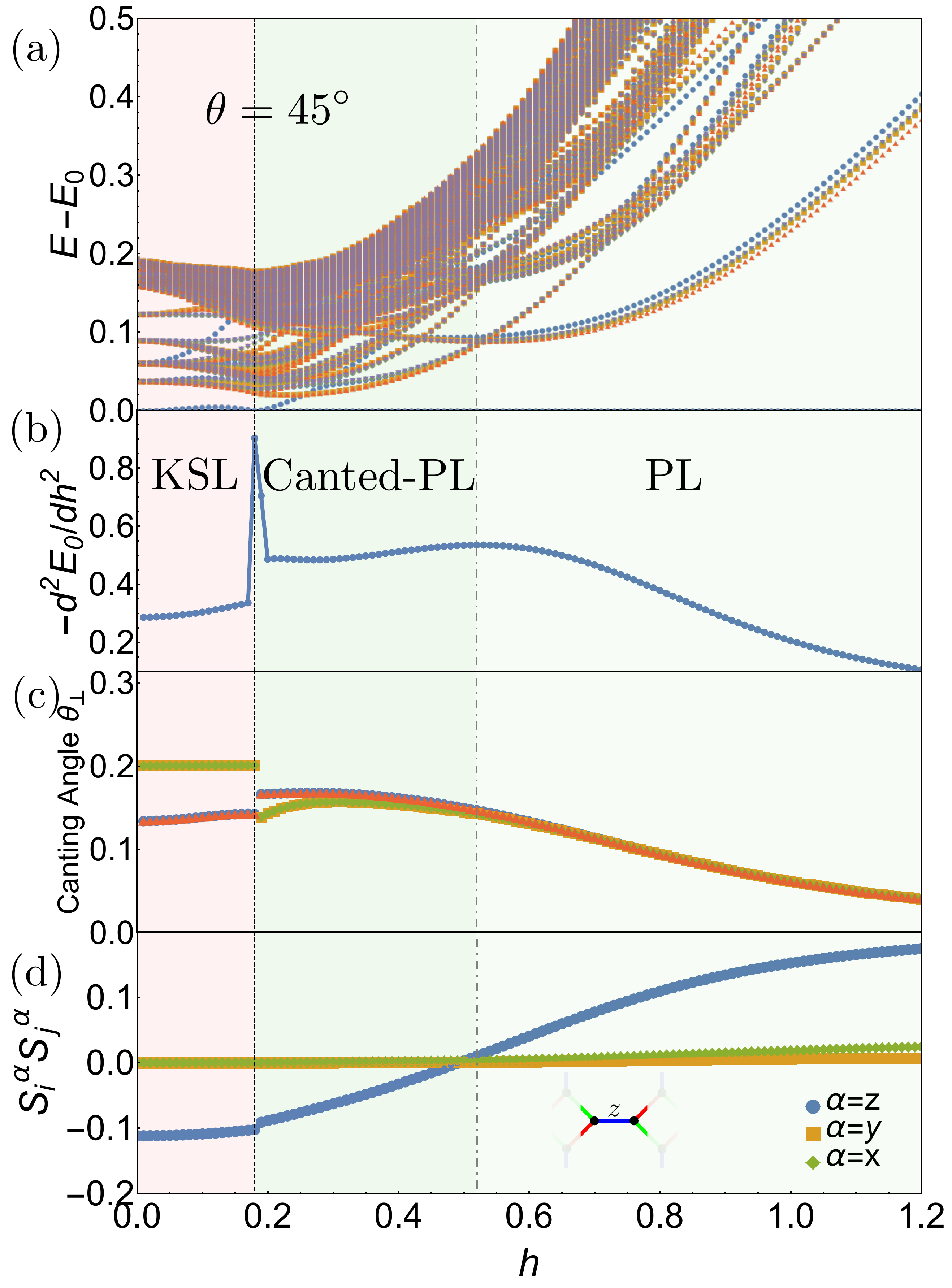}
\caption{\textbf{Phase diagram at $\theta=45^\circ$ for the AFM Kitaev model on the SO lattice} with (a) the energy spectrum (different colored points correspond to different momentum sectors), (b) the second derivative of the ground state energy, (c) canting angle $\theta_\perp$ of the local moments relative to the applied field (the 4 different markers correspond to the 4 sites in the unit cell) and (d) NN spin-spin correlations on the $z$-bonds, which connect the square plaquettes. }
\label{fig:SO_45PD}
\end{figure}

The energy spectrum for a $\theta = 45^\circ$ cut can be seen in Fig.~\ref{fig:SO_45PD}(a). There is a clear transition out of the KSL at a field of $h_c^{\text{KSL}} = 0.18$. Beyond this there is an extremely broad shoulder in the second derivative of the ground state energy, shown in Fig.~\ref{fig:SO_45PD}(b) (and associated broad dip in the fidelity, see Appendix \ref{app:gs_fidelity}). The absence of any real peak suggests that there is no phase transition. Rather, we identify the feature, centered at $h =0.52$, as a crossover between the high-field polarized regime and a canted-polarized regime at intermediate fields, $0.18 < h < 0.52$.
Fig.~\ref{fig:SO_45PD}(c) shows that the canting angle between the local moments and the applied field begins to saturate just below the crossover. Fig.~\ref{fig:SO_45PD}(d) demonstrates that the NN spin-spin correlations on the $z$-bonds, which connect squares, turn from AFM to FM at the crossover, the same behavior observed at the crossover in the DH lattice for bonds connecting triangles. 

\begin{figure}
\includegraphics[width=\columnwidth]{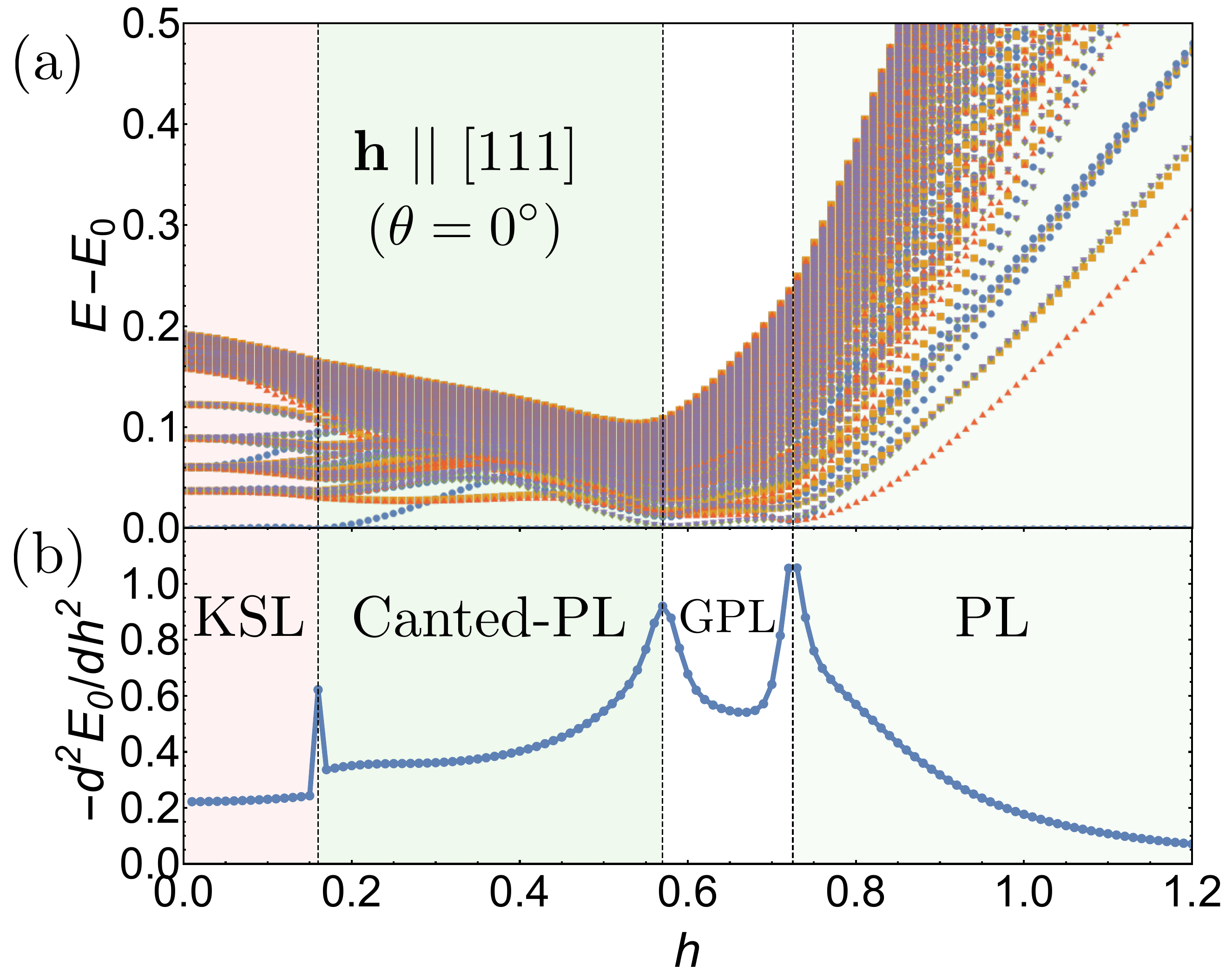}
\caption{\textbf{Phase diagram in a $[111]$ field for the AFM Kitaev model on the SO lattice} with (a) the energy spectrum (different colored points correspond to different momentum sectors) and (b) second derivative of the ground state energy. }
\label{fig:SO_ED_111PD}
\end{figure}

As the applied field is rotated toward the $[111]$ direction the broad shoulder seen in Fig.~\ref{fig:SO_45PD}(b) begins to very slowly sharpen before splitting into two clear distinct peaks for tilting angles $\theta \lesssim 7^\circ$. This splitting leads to the emergence of a new intermediate phase, as evidenced in the energy spectrum and $d^2E_0/dh^2$ results from ED shown in Fig.~\ref{fig:SO_ED_111PD}. Approaching this new phase from either side we see that the energy spectrum is driven down to low energies, resulting in an almost continuous spectrum above the ground state, suggesting that it harbors gapless degrees of freedom  
\footnote{This accumulation of low-energy states is reminiscent of the in-field behavior of the AFM KSL on the honeycomb lattice \cite{hickey_emergence_2019,hickey_field-driven_2020}.}. 
The clear peaks in $d^2E_0/dh^2$ indicate that, unlike the $45^\circ$ cut, there are true phase transitions at intermediate fields. The phase diagram for a $[111]$ field thus consists of the KSL in the region $h<h_c^{\text{KSL}} = 0.16$, the canted-PL regime in the region $0.16<h<0.57$, a new gapless (GPL) intermediate phase for $0.57<h<0.73$ and then the PL phase for fields $h>0.73$. As discussed above, it is important to keep in mind that the canted-PL regime can be smoothly connected to the high-field PL phase via a crossover once the field is tilted $\theta \gtrsim 7^\circ$ away from $[111]$. 

\begin{figure}
\includegraphics[width=\columnwidth]{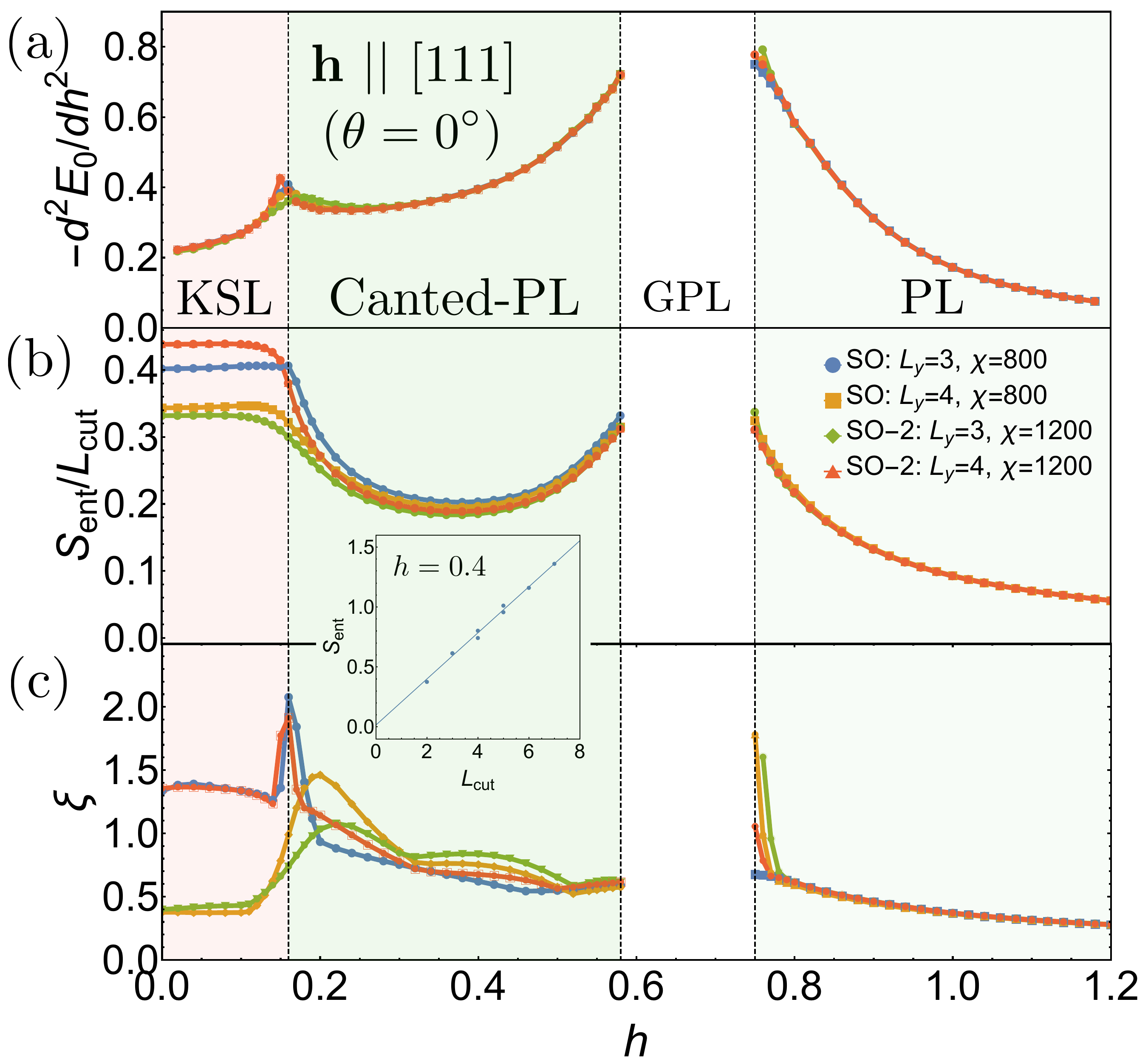}
\caption{\textbf{iDMRG results for the AFM Kitaev model on the SO lattice} showing (a) the second derivative of the ground state energy, (b) entanglement entropy $S_{\text{ent}}/L_{\text{cut}}$ and (c) the correlation length. The inset shows the scaling of $S_\text{ent}$ against $L_{\text{cut}}$ at a field value $h=0.4$ within the intermediate canted-PL regime. The linear fit is consistent with a zero intercept, indicating the absence of any topological entanglement entropy. Data within the region $0.57<h<0.73$ is not converged and thus not shown.}
\label{fig:SO_DMRG_111PD}
\end{figure}

The appearance of a new intermediate phase for a field along the $[111]$ direction can also be seen in the iDMRG data of Fig.~\ref{fig:SO_DMRG_111PD}. It is marked by an isolated region, $0.58<h<0.75$, in which the numerical results do not converge within the cylinder widths and bond dimensions studied here, indicating that the ground state within this region either harbors significant entanglement or cannot be easily captured by iDMRG. On the other hand the KSL phase has the same critical field, $h_c^{\text{KSL}} = 0.16$, as in the ED data of Fig.~\ref{fig:SO_ED_111PD}. The topologically trivial nature of the intermediate canted-PL regime is confirmed by the scaling of the entanglement entropy, shown in the inset of Fig.~\ref{fig:SO_DMRG_111PD}. At $h=0.4$, the data is fit to the form $S_{\text{ent}} = \alpha L_{\text{cut}} - \gamma$, with the resulting fit giving $\gamma = -0.02 \pm 0.03$. This is consistent with $\gamma=0$, as expected for a topologically trivial ground state.


\section{Discussion} 
\label{sec:Disc}

The multitude of detailed results presented in the preceding sections for the field-driven physics of two specific incarnations of Kitaev spin liquids
mandates that we close with a discussion of what we expect to be {\em generic} field-driven phenomena.
In doing, so we will focus on three key aspects: 
(i) the disparity of critical field strengths for KSLs arising from the sign of the Kitaev couplings,
and, concentrating on the intermediate field regime for AFM couplings,
(ii) the enhancement of spin canting, the role of lattice symmetries and the important exception of the honeycomb model, as well as
(iii) the emergence of proximate spin liquid behavior in finite-temperature and finite-energy observables.

\subsection{Critical Field Strengths}

One of the most striking features, 
which we expect to play out for all KSLs independent of their topological features and underlying lattice geometry, 
is a strong disparity in critical field strengths with FM Kitaev models transitioning out of the KSL at $h_c^{\text{KSL}} \sim \mathcal{O}(K/100)$,
versus a considerably larger value $h_c^{\text{KSL}} \sim \mathcal{O}(K/10)$ for AFM models.
The microscopic origin of this disparity can be traced back to the nature of local spin-spin correlations in the KSL. 
In the FM models these spin-spin correlations at zero-field are already of the same sign as the polarized state, whereas this is not true for the AFM models. For AFM couplings, the field needs to first flip the sign of the spin-spin correlations, from AFM to FM, and then polarize the local moments. 
This mechanism was first identified for the honeycomb Kitaev model \cite{liang_intermediate_2018,Nasu2018,hickey_emergence_2019} and subsequently discussed for a variety of 2d/3d lattice geometries on a Majorana mean-field level \cite{Berke2020}. The disparity in critical fields is clearly evident in the numerical calculations of this manuscript, with quantitative values for the critical field strengths summarized in Table \ref{tab:KSLFields}.

This disparity in critical fields is also consistent with the {\em classical} limit of the Kitaev model, as well as our semi-classical LSWT analysis. In the classical limit, the zero-field Kitaev model is a classical Coulomb spin liquid \cite{Sela2014order} at zero temperature, 
with a large manifold of degenerate spin configurations \cite{Chandra2010classical}.
For FM couplings, the polarized state is contained within this manifold and so can be immediately selected by the application of an infinitesimal field. On the other hand, for AFM couplings, the polarized state, with its FM spin-spin correlations, clearly cannot lie within the classical manifold of degenerate states. An infinitesimal field thus does not lead to immediate polarization but rather it is only reached at a finite field $h=1$ (see also Appendix \ref{app:classical_kitaev}).   

The expanded stability of the KSL in AFM systems makes the search for materials with AFM Kitaev interactions highly desirable, see also our outlook at the end of this Section. 

\begin{table}[t]
	\caption{\textbf{Overview of KSL critical fields} for a field along the $[111]$ direction, showing the order of magnitude difference between the FM and AFM models. All values are taken from ED calculations on $N=24$ site clusters, values given in brackets correspond to iDMRG results.}
	\label{tab:KSLFields}
	\begin{tabular}{l|c|c}
		Lattice & FM $h_c^{\text{KSL}}$ & AFM $h_c^{\text{KSL}}$ \\
		\hline
		Honeycomb & \,\,\, 0.025 (0.024)  \,\,\, & \,\,\,\, 0.38 (0.38) \,\,\,\, \\
		Decorated-Honeycomb  \quad \, & 0.024& 0.15 (0.10)  \\
		Square-Octagon & 0.044 & 0.16 (0.16)  \\	
	\end{tabular}
\end{table}

\subsection{Symmetry-allowed Canting}
\label{subsec:SymAllCant}
Upon closer inspection of the magnetism playing out at intermediate field strengths for the AFM Kitaev model, we have seen that the DH lattice exhibits an intermediate, canted-PL regime before a crossover into the high-field PL phase. At first, this result might seem somewhat surprising as the honeycomb model, which shares many of the same features as the DH model, instead exhibits a gapless intermediate QSL phase before a phase transition to the high-field PL phase. Both lattices share the same symmetries (the DH lattice can be constructed by simply replacing the sites of the honeycomb lattice by triangles of sites), and both share the same low-field Ising anyon topological order. Why does a magnetic field induce canting in one case and a QSL in the other?   

One crucial difference between the honeycomb and DH models is that the latter, due to its larger unit cell, admits a canting pattern, i.e.~finite $m_i^{\perp}$, that does not break any symmetry of the Hamiltonian in the presence of a $[111]$ magnetic field. It is thus possible, as the field is decreased, for the spins to smoothly cant away from the field direction and lower their energy with respect to the Kitaev interaction term. On the honeycomb lattice, it is not possible to generate canting, a finite $m_i^{\perp}$, in the presence of a $[111]$ field without breaking a symmetry \footnote{It is also true that in-plane fields parallel to the bonds, e.g.~$[1\bar{1}0]$, $[10\bar{1}]$ and $[0\bar{1}1]$, do not allow for canting without symmetry breaking. However, unlike a $[111]$ field, such in-plane fields already explicitly break the rotational symmetry of the model.}. As the field is decreased the spins are thus unable to cant away from the field direction without triggering a phase transition to a symmetry-broken state. In fact, the honeycomb lattice is special amongst the 2d Kitaev models in that it possesses such a field direction for which canting is not allowed, all other models admit some form of symmetry-allowed canting pattern regardless of the field direction (see Appendix \ref{app:canting} for more details). We note that there are instances in extended Kitaev models in which symmetries are spontaneously broken, and subsequent canting occurs, e.g.~in extended models with an additional symmetric off-diagonal $\Gamma$ exchange and a $[111]$ field \cite{Chern_magnetic_2020,Lee_magnetic_2020,Gohlke_emergence_2020}.

We are thus lead to speculate that the generic behavior of the 2d AFM Kitaev model in the presence of a field consists of a single phase transition, at $h_c^{\text{KSL}} \sim \mathcal{O}(K/10)$, from the KSL to a gapped phase with significant canted moments, followed by a crossover, at $h \sim \mathcal{O}(K)$, at which the canting rapidly decreases, all spin-spin correlations become FM and the high-field partially polarized regime is reached. Though this behavior is likely generic, there are exceptions, for special lattices or field directions. 
With regard to two-dimensional lattice geometries, the one (and probably most relevant) exception would be the honeycomb lattice, which cannot support canted moments without spontaneously breaking a symmetry for a $[111]$ field. Instead, it exhibits a gapless intermediate QSL phase. Note though that for generic field directions, i.e.~away from $[111]$, the honeycomb also allows for canted moments as the field itself will already break symmetries. Nevertheless, the gapless QSL behavior is observed to persist. In the case of special field directions, we have seen that the SO lattice offers an exception, as for fields very close to $[111]$ there is an additional intermediate phase that emerges.     
We comment on three-dimensional lattice geometries in the outlook Section.

\subsection{Proximate Spin Liquid Behavior}

Having established the formation of enhanced canted moments in the intermediate field regime as the expected generic scenario for AFM 2d Kitaev models -- with the notable exception of the honeycomb Kitaev model, one might conclude that no spin liquid physics is to be anticipated beyond the critical field strength $h_c^{\text{KSL}}$.
This, however, is somewhat premature since remnants of QSL behavior might still be found at finite temperatures and energies, {\em despite}
the trivial nature of the ground state. 
As an example, signatures of fractionalization have been observed in the finite-energy dynamics of 1d spin chain compounds \cite{Lake2013Multispinon}, even though the ground state is ultimately a 3d long-range ordered state (due to the always present interchain couplings). In the context of the Kitaev materials, this concept has been used to describe for example the similarities between the high-energy continuum observed in neutron scattering \cite{Banerjee2016proximate,Banerjee2016neutron} or resonant inelastic X-ray scattering (RIXS) \cite{Revelli2019fingerprints} experiments and the continuum that arises in the Kitaev model and its extensions \cite{Knolle2014dynamics,Knolle2015dynamics,Gohlke_dynamics_2017}. Such similarities have led to the notion of a ``proximate spin liquid'' regime \cite{Banerjee2016proximate,Banerjee2016neutron}. 

\begin{figure}[t]
\includegraphics[width=\columnwidth]{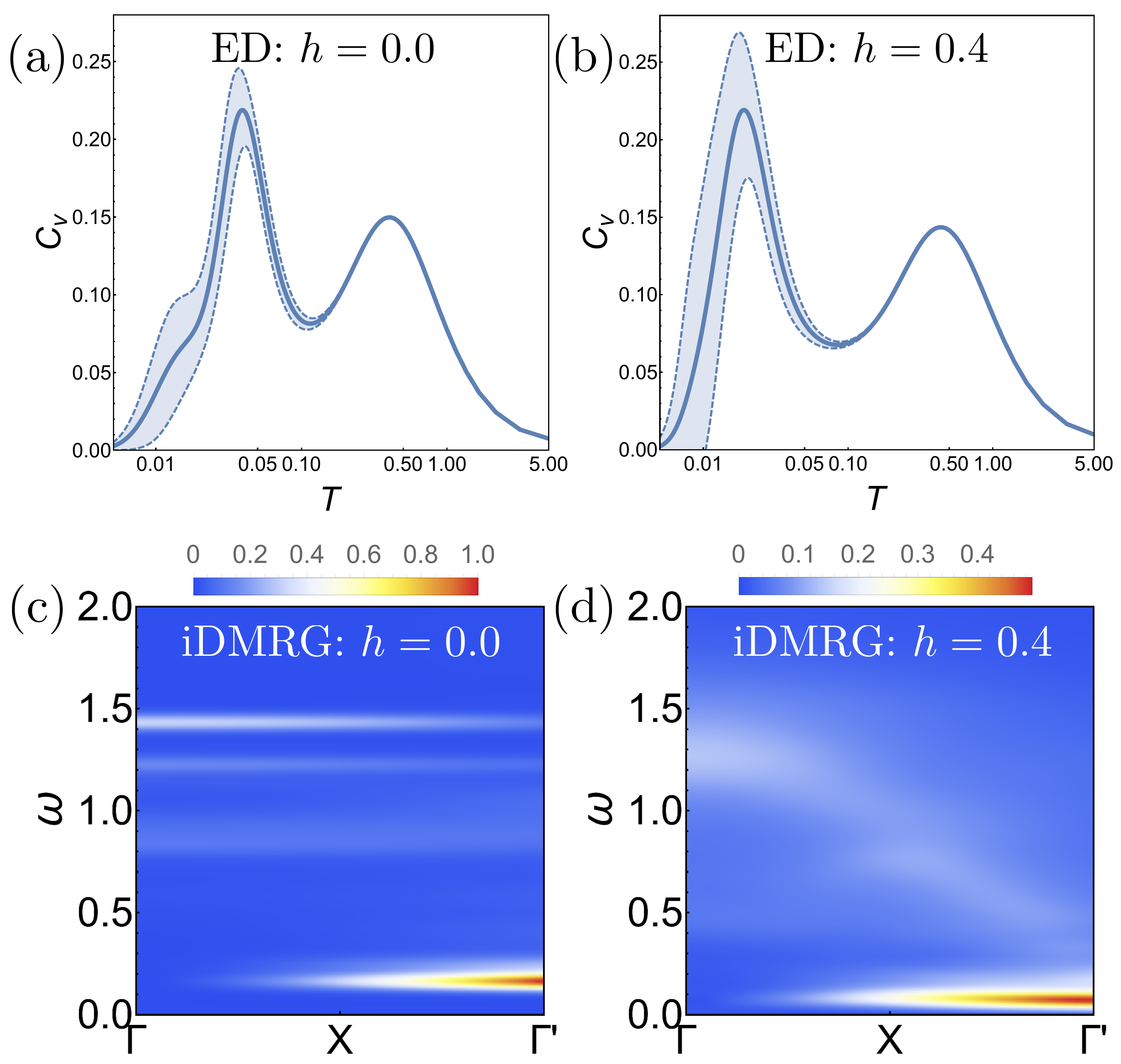}
\caption{\textbf{Comparison of spin liquid and proximate spin liquid behavior.} 
		Shown are the specific heat, (a), (b), and dynamical structure factor, (c), (d),
		for parameters deep in the KSL regime, $h=0.0$, (a) and (c), and the canted-polarized regime, $h=0.4$, (b) and (d).
		All data is for the SO lattice and a field along the $[111]$ direction.} 
\label{fig:SO_PSL}
\end{figure}

At finite temperature, one of the key characteristics of the Kitaev model is the double peak structure of its specific heat \cite{Nasu2014vaporization},
as illustrated in the upper left panel of Fig.~\ref{fig:SO_PSL} for the SO model in zero field. The two peaks indicate subsequent crossovers from the high-temperature paramagnetic regime into a fractionalized, flux-disordered intermediate regime to the true (flux-ordered) KSL as temperature is lowered.
Under a magnetic field, even though the KSL ground state is destroyed at $h_c^{\text{KSL}}$, this double peak structure is found to persist to considerably higher fields. 
The upper right panel in Fig.~\ref{fig:SO_PSL} shows such a specific heat trace midway within the canted-PL regime for a field $h=0.4$ along the $[111]$ direction, more than a factor of 2 above the critical field strength of $h_c^{\text{KSL}} = 0.16$ for the SO model at hand, but still almost identical to the zero-field specific heat trace.
A similar picture is found when going to finite energy, e.g. by looking at the dynamical structure factor of the Kitaev model. 
As plotted in the lower left panel of Fig.~\ref{fig:SO_PSL} the dynamical structure factor of the SO KSL exhibits a sharp gap equal to the two-vison gap, as $S_i^\alpha$ flips the value of $W_p$ on two adjacent plaquettes creating two gapped visons (see Appendix \ref{app:visons} for more details). The static nature of the vison excitations results in a flat dispersionless band. Above this sharp feature there is a broad continuum of fractionalized excitations, a key signature of the non-trivial nature of the Kitaev model. 
Interestingly, if we look again at $h=0.4$ midway within the SO canted-PL regime, the dynamics looks strikingly similar to that of the pure Kitaev model. The dynamical structure factor is shown in Fig.~\ref{fig:SO_PSL}(d). There is again a sharp gap above which lies a flat dispersionless band, followed by a broad continuum, albeit slightly smeared out compared to the zero-field limit.  
It is quite remarkable that 
the SO canted-PL phase in the intermediate-field regime shares, at finite temperatures and energies, so many of the same characteristics features of the pure Kitaev model, despite having a topologically trivial ground state (as clearly demonstrated in the scaling of $S_\text{ent}$ in the inset of Fig.~\ref{fig:SO_DMRG_111PD}). It is as such a rather clean example of what is considered proximate spin liquid behavior. 

\begin{figure}
\includegraphics[width=\columnwidth]{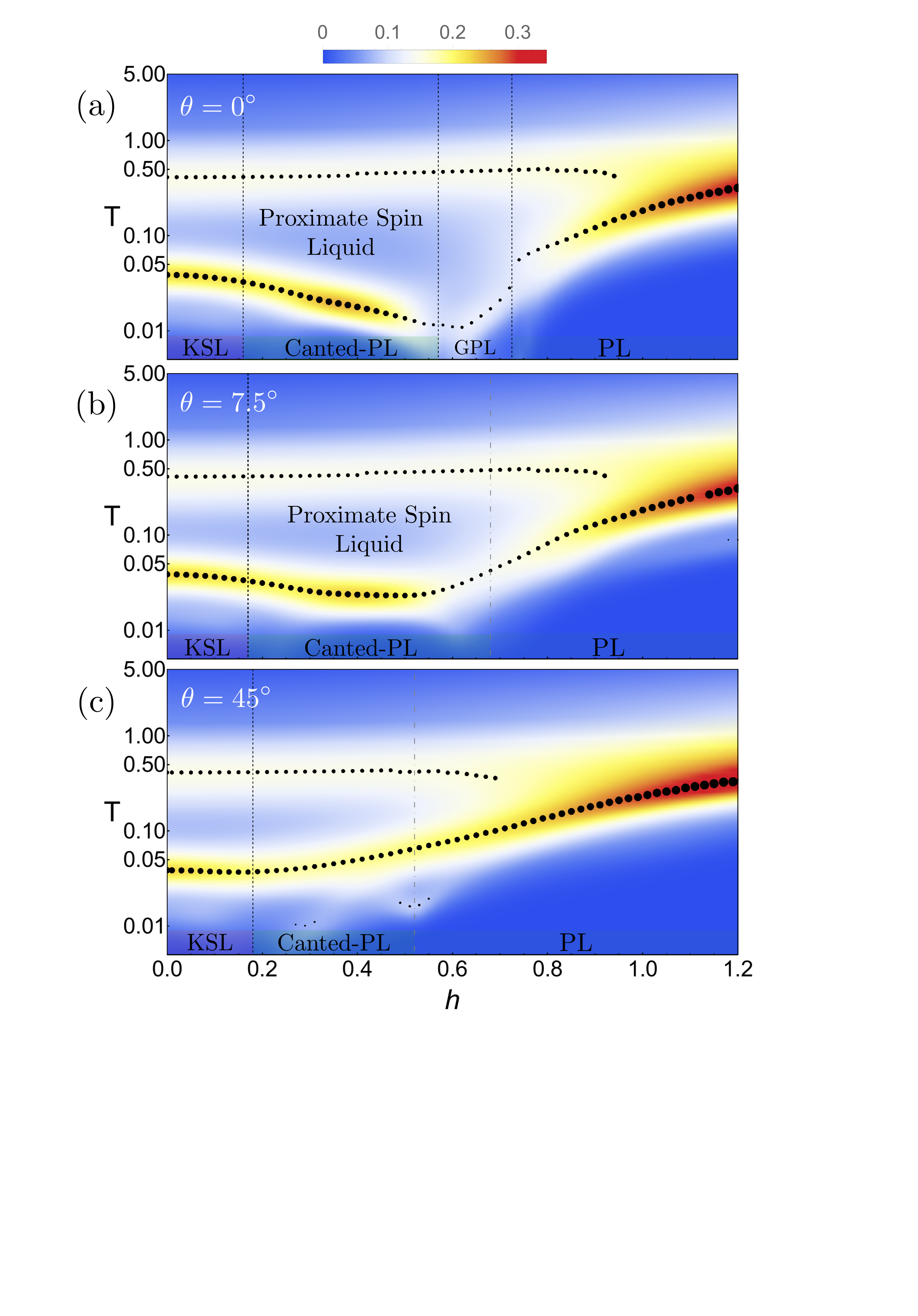}
\caption{\textbf{Thermodynamic signature of proximate spin liquid behavior.} 
		The color-coded specific heat for the SO lattice at different tilt angles exhibits a distinct two-peak structure
		in a broad field range beyond the actual KSL phase.}
\label{fig:SO_Cv_Tilted}
\end{figure}

Eventually, for sufficiently strong field strengths, the proximate spin liquid regime disappears. In the specific heat data, for instance, this can be seen as the two peaks eventually merge into a single peak. This is shown for a variation of different tilt angles in Fig.~\ref{fig:SO_Cv_Tilted}, again for the AFM SO model.
For $\theta=45^\circ$ (lowest panel), we see the principle signature of an extended proximate spin liquid regime with the lower of the two peaks in the specific heat slowly shifting upwards before merging with the upper peak, similar to what is shown in Fig.~\ref{fig:YK_Cv} for the DH lattice.
Going to smaller tilt angles, the $\theta=7.5^\circ$ cut in Fig.~\ref{fig:SO_Cv_Tilted}(b), the proximate spin liquid regime is even more pronounced 
as the lower peak slightly {\em decreases} on entering the canted-PL regime and retains this position over a wide field range. For $\theta=0^\circ$, i.e.~a field along the [111] direction, the lower peak is pushed down to even lower temperatures and loses its structure on entry to the gapless phase. As a result, the specific heat within the canted-PL regime close to $[111]$ looks almost identical to that of the pure Kitaev model itself -- this is what we shown in the comparison of Fig.~\ref{fig:SO_PSL}(a), taken at $h=0$, and Fig.~\ref{fig:SO_PSL}(b), taken at $h=0.4$ midway within the canted-PL regime.   

Notably, such a proximate spin liquid regime is found in both the  AFM and FM models, though it persists over a much wider field range in the AFM model. In fact, for the DH lattice, the ratio of the phase boundaries $h_{sp}/ h_c^{\text{KSL}}$ (where $h_{sp}$ is the field strength at which the two peaks in the specific heat merge into a single peak) is roughly the same for both cases. This suggests that the extended double peak regime in the AFM model is, in fact, due to the extended stability of its KSL phase. That is, the more stable KSL phase also induces a wider proximate spin liquid regime.  

\subsection{Outlook}

Taking a look ahead, our study reemphasizes the interest in AFM Kitaev materials. These systems not only exhibit KSL phases which are considerably more stable in the presence of a magnetic field. They also give rise to an extended intermediate-field strength regime that -- despite the generic formation of a conventional ground state with canted magnetic moments for most lattice geometries -- exhibits proximate spin liquid behavior in its finite-temperature and finite-energy observables. A route towards the experimental exploration of such AFM Kitaev materials has been laid out theoretically, identifying $4f$ materials \cite{Jang_antiferromagnetic_2019,sugita_antiferromagnetic_2019} as well as spin-1 Kitaev materials \cite{Stavropoulos_microscopic_2019,hickey_field-driven_2020} as alternatives 
to the current family of $4d/5d$ spin-orbit entangled $j=1/2$ Mott insulators that have predominantly FM Kitaev couplings
\cite{Rau2016,Winter2016challenges,Trebst2017,motome_materials_2020}. 

A particularly intriguing direction to pursue is the exploration of 3d Kitaev materials, such as the hyperhoneycomb systems $\beta$-LiIr$_2$O$_3$ \cite{Takayama2015hyperhoneycomb} and $\gamma$-LiIr$_2$O$_3$ \cite{Modic2014realization}, but with dominant AFM couplings. While such systems might have an even stronger tendency to exhibit magnetically ordered ground states compared to their 2d counterparts, the prospect of experimentally realizing proximate spin liquid physics in a 3d system is tantalizing. It would provide us with a setting to explore signatures of fractionalization that is quite different from other 3d spin liquid materials, pursued, e.g., in the search for a Coulombic quantum liquid \cite{Savary_coulombic_2012} in the spin-1/2 rare earth pyrochlore systems \cite{Rau_2019_frustrated}. 
Particularly interesting lattice geometries to realize in candidate materials might be 
the hyperhoneycomb and its higher harmonics (see Appendix \ref{app:canting}), 
which similar to the honeycomb lattice in two spatial dimensions obstruct spin canting for certain field directions
due to their underlying symmetries.


\acknowledgments 
We thank L.~Savary and M.~Vojta for useful discussions. This work was supported by the Deutsche Forschungsgemeinschaft (DFG, German Research Foundation) -- Projektnummern 277101999 and 277146847 -- TRR 183 (project A04) and SFB1238 (project C02). This research was supported in part by the National Science Foundation under Grant No. NSF PHY-1748958. Part of the numerical simulations were performed on the CHEOPS cluster at RRZK Cologne and the JUWELS cluster at FZ J{\"u}lich. 
M.G. acknowledges support by the Theory of Quantum Matter Unit of the Okinawa Institute of Science and Technology Graduate University (OIST) and by the Scientific Computing section of the Research Support Division at OIST for providing the HPC ressources.



%


\appendix

\section{iDMRG Geometries}
\label{app:dmrg_geometries}

\begin{figure}
\includegraphics[width=\columnwidth]{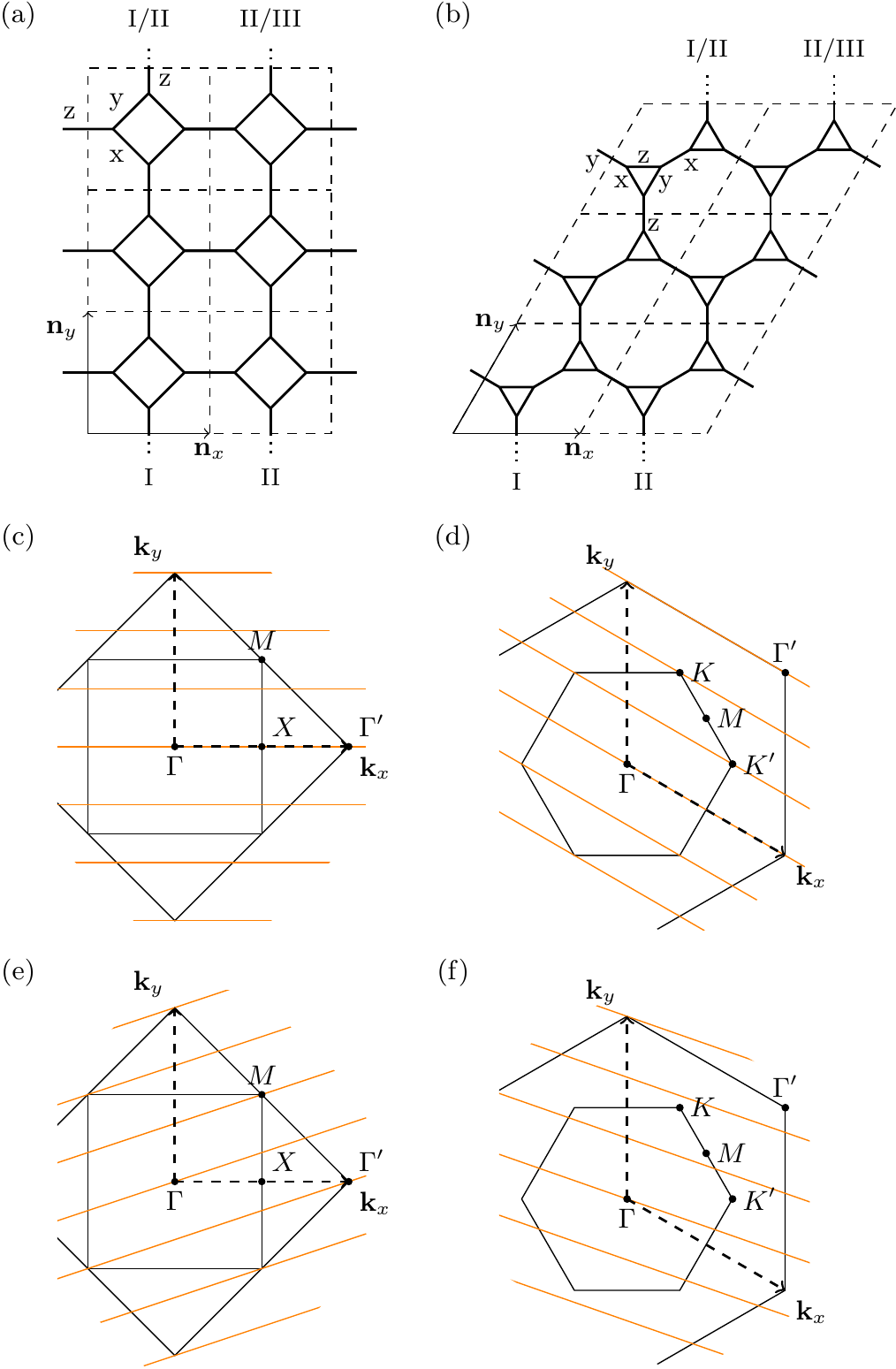}
\caption{\textbf{iDMRG cylinder geometries} and reciprocal space of the SO (a,c,e) and DH (b,d,f) models. The roman numbers denote the way the bonds are connected across the boundary. Two geometries are used: (i) a \emph{ringlike} geometry denoted as SO or DH, where the top and bottom bond belong to the same row, and (ii) a geometry with \emph{shifted} boundary condition, denoted as SO-2 and DH-2.
Both geometries result in different cuts of accessible momenta, $\mathbf{k}$, in reciprocal space. These cuts are either parallel to the $\mathbf{k}_x$ vector for SO (c) or DH (d), respectively, or tilted for SO-2 (e) or DH-2 (f).}
\label{fig:dmrg_geometries}
\end{figure}

In order to be able to use iDMRG, we wrap the two-dimensional DH and SO lattices on a cylinder and wind the one-dimensional matrix product state structure around the cylinder. Due to the cylindrical geometry, only discrete steps of the wave vectors $\mathbf{k}_y/L_y$ are available, where $\mathbf{k}_y$ is the reciprocal lattice vector corresponding to $\mathbf{n}_y$ and $L_y$ is the number of unit cells along the circumference. This results in lines of accessible momenta in the reciprocal space.
The way in which the lower and the upper boundaries are connected determines 
the orientation of these lines, while the circumference determines the spacing between them.

Here, we employ two different ways of connecting the upper and the lower bond across the boundary, \emph{ringlike} which we denote as SO and DH, and \emph{shifted} which we denote as SO-2 and DH-2. 
See Fig.~\ref{fig:dmrg_geometries} for an illustration of each geometry and their corresponding accessible momenta. We restrict ourselves to circumferences $L_y = 3$ and $4$ when computing the ground state due to an exponentially with $L_y$ increasing computational cost.
The time evolution is performed on the \emph{ringlike} geometries, SO and DH, with $L_2 = 3$.

\section{Classical Kitaev Model in Field}
\label{app:classical_kitaev}

In its classical limit, the Kitaev honeycomb model is known to evade any finite-temperature ordering transition and to exhibit a classical spin liquid ground state that can be framed as a Coulomb phase \cite{Baskaran2008,Chandra2010classical,Sela2014order}. On a more technical level, the exchange frustration -- arising from the competing bond-directional interactions also on the classical level -- leads to an extensive manifold of classical spin configurations with the same minimum ground state energy, that distinguish themselves from the high-temperature paramagnet by the formation of local dimers of spins pointing parallel along one of three principle spin axes. This dimer formation allows to map the manifold of degenerate spin configurations to the manifold of dimer coverings of the underlying lattice. Since the origin of this degeneracy-inducing mechanism plays out at the level of a single tricoordinated vertex, we expect it to play out for all classical Kitaev models, independent of their tricoordinated lattice geometry and spatial dimensionality. Though the details of the classical dimer model might still subtly differ depending on the lattice, in all cases a classical spin liquid emerges as the classical ground state.   

In the presence of a $[111]$ magnetic field, the FM Kitaev models are immediately polarized as the classical fully polarized state is contained within its manifold of ground states. As a result, LSWT about the polarized state can be easily carried out at any finite field. On the other hand, the AFM Kitaev models exhibit extremely non-trivial behavior in field. We will not go into details of this physics here but instead we merely point out that the fully polarized state is only stable for fields $h>1$ (assuming classical spins normalized to $|\vec{S}| = 1/2$). This means that LSWT can only be reliably carried out for $h>1$. 

\begin{figure}
\includegraphics[width=\columnwidth]{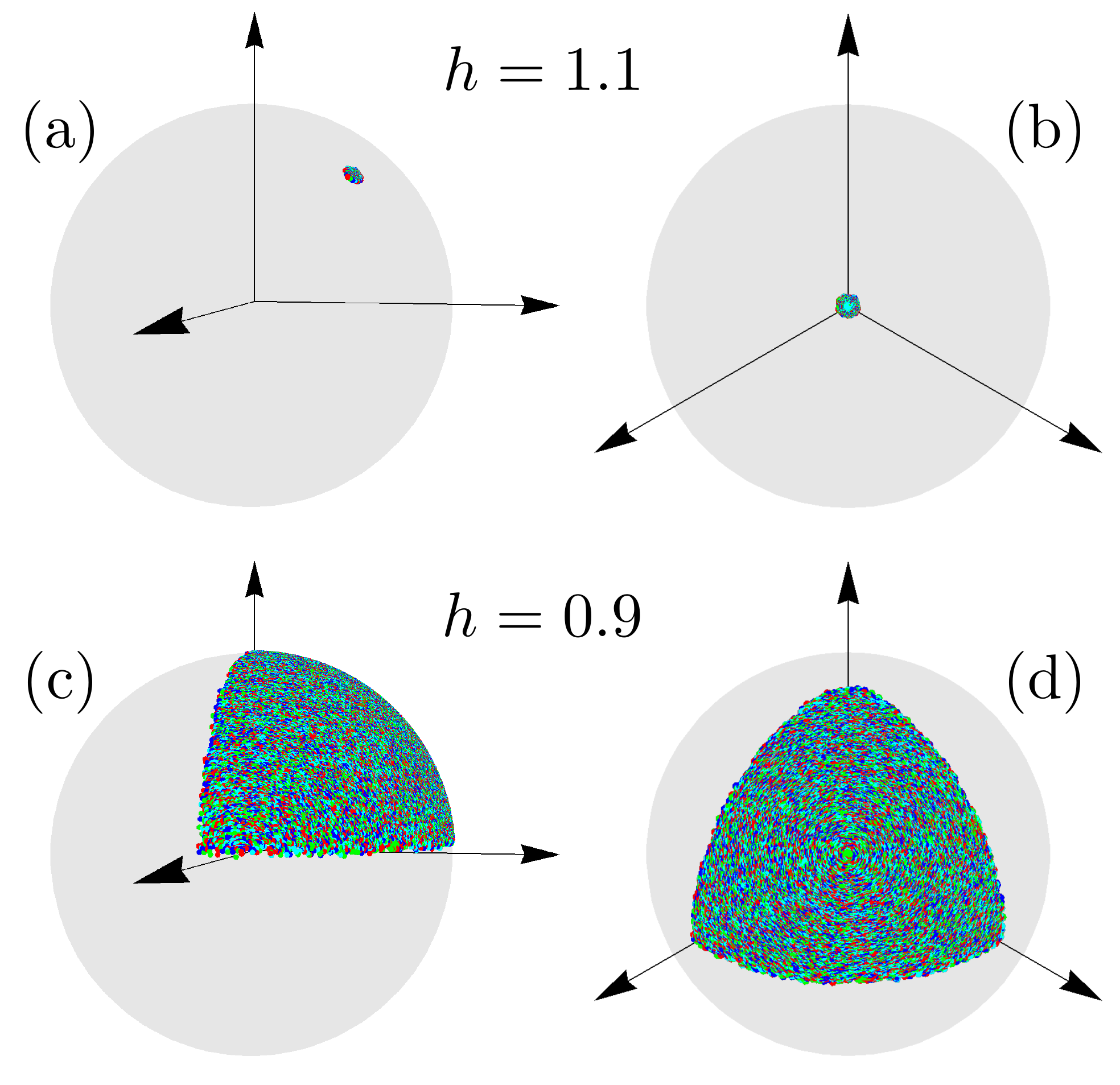}
\caption{\textbf{Common origin plots} for the classical AFM Kitaev model on the SO lattice.}
\label{fig:Classical}
\end{figure}

In Fig.~\ref{fig:Classical} we show a series of common origin plots, in which all of the spins in a classical spin configuration are plotted with the same base point, for the classical AFM Kitaev model on the SO lattice as an example. The configurations are generated via simulated annealing of lattices of $96 \times 96$ unit cells, resulting in $36,864$ spins for the SO lattice, and taken at a final temperature $T=10^{-5}$. For $h=1.1$ we see in Fig.~\ref{fig:Classical} that the ground state is clearly the polarized state, with all of the spins point along the $[111]$ field direction. On the other hand, for a field $h=0.9$, we see in Fig.~\ref{fig:Classical} that the ground state spin configuration is disordered, albeit constrained to a specific region of the sphere (a region which varies as $h$ decreases). As there is no well-defined magnetic unit cell, it is thus not possible to carry out conventional LSWT for such disordered configurations. Interestingly we note that the difference in the energy per site between the disordered configurations generated via simulated annealing and the translationally invariant configuration produced by minimizing the energy of a single unit cell is $\sim 10^{-5}$ (with the analytic expression for the energy per site from minimizing a single unit cell given by $E_0 = [ K(h^2 - 2 S^2) - 2 h^2 ]/4$ for $h<1$). Similar results hold for the honeycomb and decorated honeycomb lattice.

\section{Magnon Dimensional Reduction within LSWT}
\label{app:dimensional_reduction}

We would like to briefly note an interesting quirk within LSWT of the polarized phase of the Kitaev model. For fields in which one or more components vanish the magnon spectrum becomes {\em flat} along certain directions in momentum space. For example, if we were to take $\vec{h}=(0,0,h_z)$, then the spins will be classically polarized purely along the $z$-axis. In this case, terms on the $z$-bonds, $S_i^z S_j^z$, will only contribute on-site terms, $b_i^\dagger b_i$ and $b_j^\dagger b_j$, within LSWT. This means that there is no hopping along the $z$-bonds, but instead only along $x$ and $y$ bonds. 

The most dramatic example of this occurs in the SO lattice. In this case, if the bond connecting squares is the $z$-bond, then a field along the $z$-axis will result in bosons purely confined within each square, hopping between squares vanishes. The resulting magnon bands are thus completely flat due to the $0$-dimensional nature of the resulting spin-wave Hamiltonian. If instead the field is chosen to point purely along either the $x$ or $y$ axis then the resulting magnon bands will be flat only along certain directions in momentum space. In other words the Hamiltonian within LSWT only contains $1$-dimensional hopping terms, along chains in the lattice. 

The reduction of the full 2-d magnon dispersion of the polarized state to a 1-d dispersion, i.e. the appearance of flat directions in momentum space, occurs for magnetic fields parallel to the cubic axes in all 2-d/3-d Kitaev models. This is related to the fact that all models admit Jordan-Wigner strings that fully cover all sites. Of course, it should be emphasized that the exact flatness is a quirk of LSWT, corrections beyond LSWT will induce interactions across all bonds, destroying the reduced dimensionality of the Hamiltonian.

\section{Ground State Fidelity}
\label{app:gs_fidelity}

The ground state fidelity is defined as $F(g) = |\avg{\Psi_0 (g) | \Psi_0 (g+\delta g)}|$ for some tuning parameter $g$. It is an order-parameter independent quantity that can signal the presence of phase transitions as $g$ is varied. In the thermodynamic limit, the associated susceptibility, $d^2F/dg^2$, exhibits characteristic scaling behavior that can further distinguish between first and second order transitions. In ED, a transition which occurs via a level crossing in the ground state results in $F(g)=0$ while a transition via mixing between the ground state and excited states results in a smooth dip in $F(g)$. Within a phase, away from any transitions, $F(g) \approx 1$.

\begin{figure}[h!]
\includegraphics[width=\columnwidth]{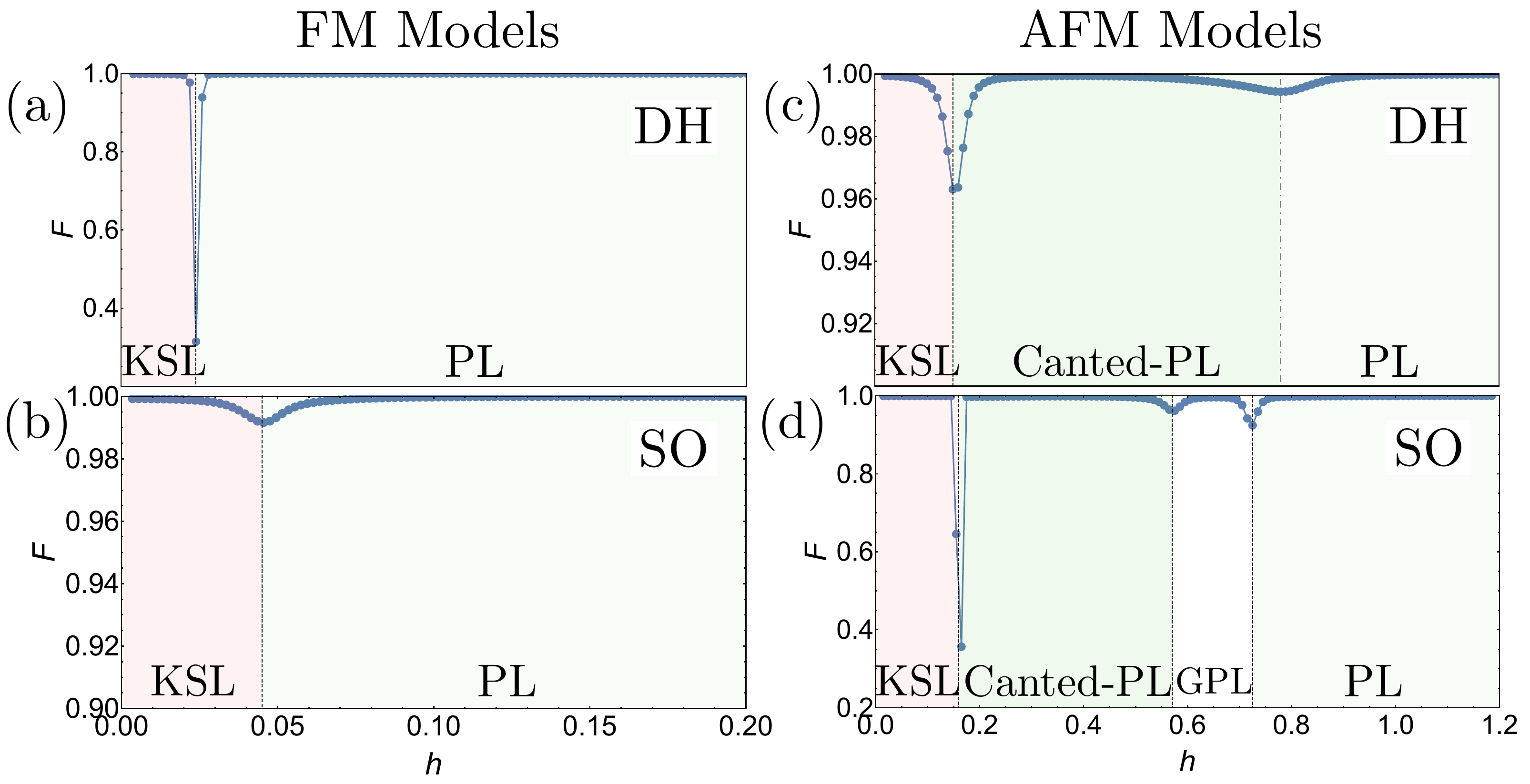}
\caption{\textbf{Fidelity traces} for the in-field phase diagrams of the DH and SO models. }
\label{fig:Fidelity}
\end{figure}

To complement the phase diagrams in the main text we show the ground state fidelity $F(h)$ for the FM Kitaev models on the DH and SO lattice, as well as for the AFM Kitaev models on the same lattices (in all cases for a $[111]$ field). The KSL transition at $h_c^{\text{KSL}}$ is clearly visible in all cases. On the other hand, the crossover from the canted-PL regime to the high-field polarized regime fro the DH lattice only produces a weak change in $F(h)$ (note the extremely fine scale on the $F(h)$ axis in Fig.~\ref{fig:Fidelity}). The intermediate gapless phase for the SO lattice is also clearly visible in the fidelity.    

\section{Vison Gaps}
\label{app:visons}

In the Kitaev model, there is a finite energy cost to flipping a plaquette $W_p$, which, in the context of the exact solution, corresponds to a finite energy cost to making a $\mathbb{Z}_2$ gauge excitation, or ``vison''. The operators $S_i^\alpha$ acting on the ground state create two visons in the plaquettes adjacent to the $\alpha$-bond. Unlike the honeycomb model, the DH and SO lattices possess two different types of plaquettes, and thus two different vison gaps $\Delta_v$ (note that here we are using $\Delta_v$ to denote the gap to creating two visons on adjacent plaquettes, not the gap to creating a single vison). The gaps can be straightforwardly calculated at zero field within the Majorana reformulation of the model. 

\begin{figure}[t]
\includegraphics[width=\columnwidth]{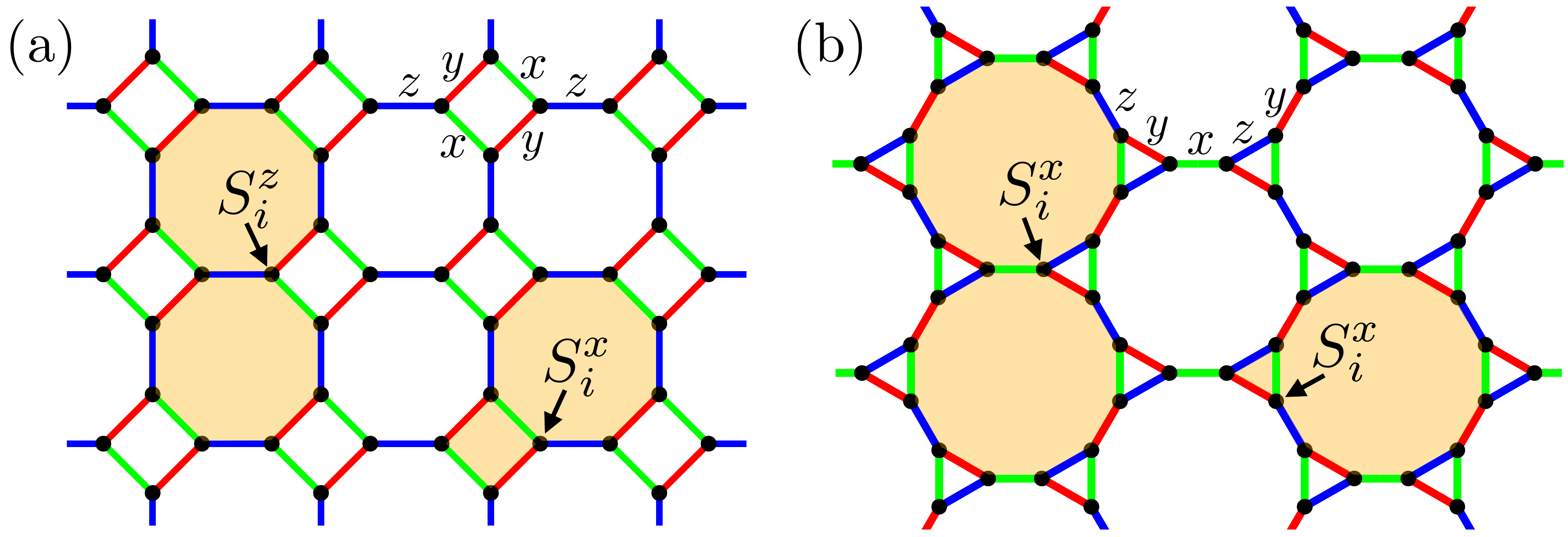}
\caption{\textbf{Two-vison excitations} on the (a) SO lattice and (b) DH lattice, created by acting $S_i^\alpha$ on the sites indicated. }
\label{fig:Visons}
\end{figure}

The different potential vison gaps are illustrated in Fig.~\ref{fig:Visons}, with (a) showing how, on the SO lattice, $S_i^z$ always flips two octagonal plaquettes whereas $S_i^x$ (or $S_i^y$) will always flip one octagonal plaquette and one square plaquette. The associated energy costs are $0.02 K$ and $0.13 K$ respectively. In (b) we see how, on the DH lattice, $S_i^x$ can either flip two dodecagonal plaquettes or one triangular plaquette and one dodecagonal plaquette, depending on the site $i$ (similarly for $S_i^y$ and $S_i^z$). Here, the energy costs for such flips are $0.07K$ and $0.03K$ respectively (note that the vison gap for two-dodecagon excitations is almost identical to that for two-hexagon excitations on the honeycomb lattice, if one replaces the triangles of the DH lattice by sites the dodecagons become the hexagons of the honeycomb lattice). 

\begin{figure}[t]
\includegraphics[width=\columnwidth]{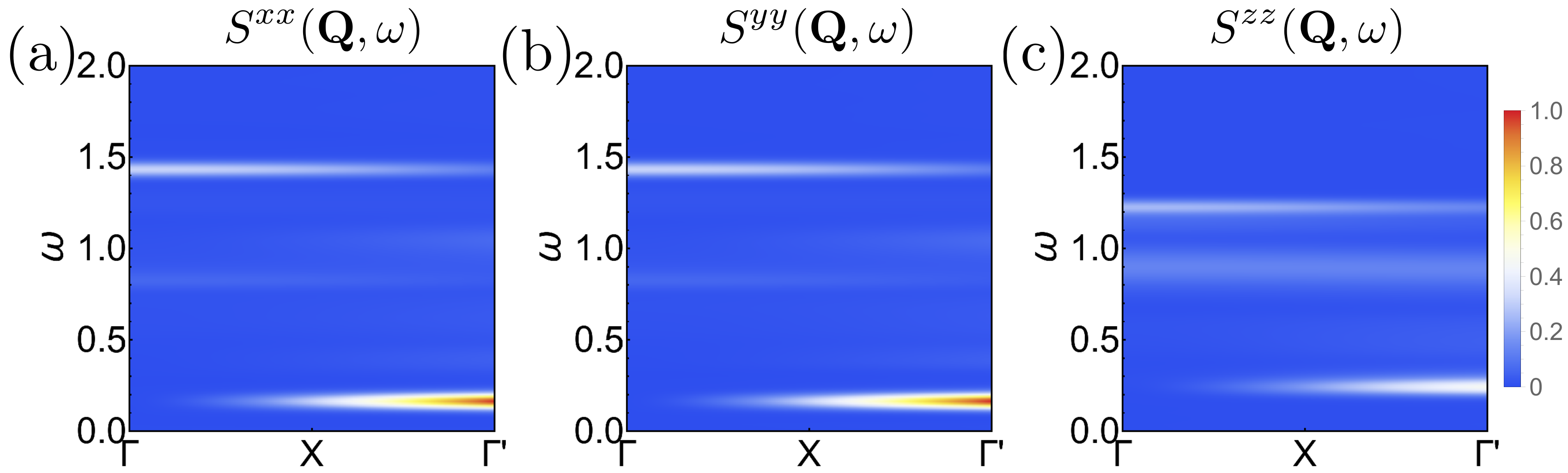}
\caption{\textbf{Spin-resolved dynamical structure factor} for the SO lattice at zero field.}
\label{fig:DSFVisons}
\end{figure}

The finite vison gap can be directly observed in the dynamical structure factor which, at zero field, has a sharp gap equal to $\Delta_v$. For the DH lattice there is no clear way to disentangle the two different vison gaps. However, owing to the unique structure of the square-octagon lattice, which has one bond distinct from the other two (the bond connecting squares), it's actually possible to separately observe both vison gaps by looking at the spin-resolved components of the dynamical structure factor. In Fig.~\ref{fig:DSFVisons} we show the diagonal components of the dynamical structure factor, $S^{\alpha\alpha} (\vec{Q},\omega)$, calculated using iDMRG. There is a clear difference between the $S^{xx}$/$S^{yy}$ and $S^{zz}$ components, with $S^{xx}$/$S^{yy}$ sensitive to the square-octagon vison gap and $S^{zz}$ sensitive to only the octagon-octagon vison gap. The specific values of the gaps are note the same as those found via the exact solution due to the cylinder geometry used in the iDMRG simulations.      

\section{Symmetry-allowed Canting Patterns}
\label{app:canting}
In this Appendix, we extend the discussion of the symmetry-allowed canting in Sec. \ref{subsec:SymAllCant}. For all lattices we discuss the symmetries of the Kitaev model and how they are affected by a [111] field. 
We also examine generalizations of the Kitaev model to three dimensional lattices, with focus on field directions along high-symmetry lines.
In the following, we write \textbf{M}$_i$ for the magnetization at unit cell position $i$.
To start, we review Kitaev's original model on the honeycomb lattice.

\subsection{Honeycomb lattice}
\begin{figure}
	\includegraphics[scale=1]{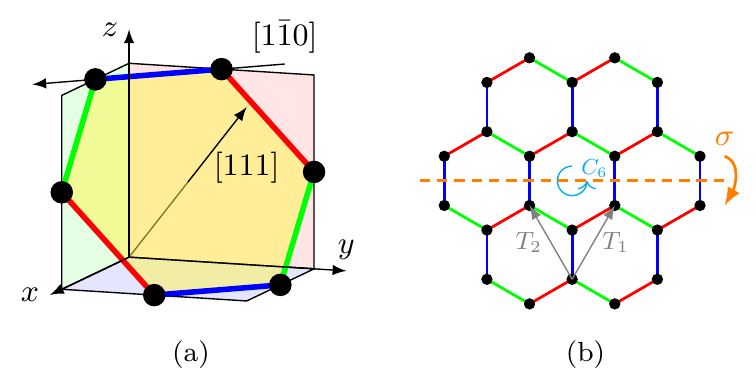}
	\caption{{\bf The honeycomb lattice} can (a) be embedded in 3d such that all symmetry operations act equally on lattice and spin. The [111] direction is perpendicular to the lattice plane. (b) Symmetries of the Kitaev model. The $C_6$ symmetry is composed of a sixfold rotation and a reflection at the lattice plane.}
	\label{fig:HCsym}
\end{figure}

Due to the strong spin-orbit interactions, the symmetries act on both spin and spatial degrees of freedom. It turns out to be especially useful to embed the lattice in 3d with the [111] (\textbf{c}) direction pointing perpendicular to the lattice plane, as visualized in Fig. \ref{fig:HCsym}(a) \cite{You2012doping}. Precisely the same orientation is found in the Iridium-based material realizations. The symmetries act then equally on spin and lattice degrees of freedom. 

The symmetry group of the Kitaev model contains an operation $C_6$ composed of a 6-fold $\textbf{c}$-axis rotation followed by a reflection across the lattice plane, and a reflection $\sigma$ across the $x=y$ plane. Together with translations $T_1,T_2$ and time reversal $\mathcal{T}$, the full symmetry group is SG $=\langle \mathcal{T}, T_1, T_2, C_6,\sigma \rangle$. The symmetries are visualized in Fig. \ref{fig:HCsym}(b). 

Upon switching on a magnetic field in [111] direction, mirror and time reversal symmetry are individually broken but their combined action and $C_6$ remain symmetries. Twofold application of $C_6$ preserves the position within a unit cell but maps $(\sigma_x,\sigma_y,\sigma_z) \to (\sigma_y,\sigma_z,\sigma_x)$. The magnetization can thus not deviate from the field direction without breaking further symmetries.

The $C_6$ symmetry is broken by any field component perpendicular to the [111] direction. It is then natural to ask whether there are special field directions that preserve some of the symmetries and whether within this subset, canting is allowed. Apart from translation symmetry and inversion, the combination $\mathcal{T}\sigma$ is preserved for fields lying in the $[11x]$ plane \cite{zou_field-induced_2019}. The magnetization might deviate from the field direction without breaking any further symmetries, as long as its $[1\bar{1}0]$ component vanishes. In contrast, no canting is allowed for fields pointing along the lattice bonds, e.g. $[1\bar{1}0]$. Such a field preserves $\sigma$ \cite{zou_field-induced_2019} and this alone would allow for a canting pattern where the $[111]$ components of \textbf{M}$_1$ and \textbf{M}$_2$ have equal magnitude but opposite signs. However, as \textbf{M}$_1$ and \textbf{M}$_2$ are also related by inversion symmetry, $\textbf{M}_1=\textbf{M}_2$ must hold for all symmetry preserving patterns. 

Any other field direction generically allows canting as all symmetries except for inversion and translation are broken.

\subsection{Decorated honeycomb lattice}
Replacing each site of the honeycomb lattice by a triangle gives rise to a new structure called decorated honeycomb lattice, shown in Fig. \ref{fig:Lattices} \cite{yao-kivelson}. The Kitaev model on this lattice has the same symmetry group as its honeycomb cousin. In contrast to the latter, only full $2\pi$ rotations map one position in a unit cell to itself. The restriction for \textbf{M} in a [111] field that emerged as a consequence of the twofold application of $C_6$ is therefore omitted. The more complicated unit cell allows the magnetization to cant away from the field direction. More precisely, $C_6$ enables a continuous manifold of configurations that can be constructed by simultaneously rotating all spins around the \textbf{c} direction by an arbitrary angle, starting in an \textit{all in all out} order, where all spins in the left (right) pointing triangles in Fig. \ref{fig:Lattices} point inwards (outwards). The symmetry $\mathcal{T}\sigma$ picks two special canting patterns of this manifold, namely the ones where the in-plane component of \textbf{M}$_i$ is perpendicular to the bond connecting site $i$ to a different triangle. These are the two configurations shown in Fig. \ref{fig:YK_ED_PD}.

\subsection{Square-octagon lattice}
The square-octagon lattice is visualized in Fig. \ref{fig:Lattices}. Its space group consists of a fourfold rotation around the axis perpendicular to the lattice and four reflection at planes perpendicular to the lattice plane, which intersect and halve the four different bonds vertically (note that there are two different $z$-bonds).
The most natural way to embed the lattice in 3d is to select $x$- and $y$-axis parallel to the respective bonds and the $z$-axis perpendicular to the lattice plane. This guarantees that a real space symmetry transformation, supported by the identical operation in spin space is a symmetry of the Kitaev model \footnote{Note that not all lattices allow for such a global choice of the coordinate system for real and spin space, for an exception see e.g. the lattice (10,3)c.}
The symmetry group of the Kitaev model is further enriched by $\mathcal{T}$ and reflection at the lattice plane $\mathcal{L}$. The latter leaves the lattice itself invariant but maps $(\sigma_x,\sigma_y,\sigma_z)\to(-\sigma_x,-\sigma_y,\sigma_z)$. Together with translations $T_1,T_2$, the full symmetry group is generated by e.g. $\langle T_1,T_2,R,M_x,\mathcal{T},\mathcal{U} \rangle$, where $M_x$ is the reflection at the plane with the normal vector $\hat{x}$.  

Unlike in the honeycomb model, a [111] field breaks the rotation symmetry. The residual symmetries are $\mathcal{L}R^2$, $\mathcal{T}M_{x\bar{y}}$ and $\mathcal{T}\mathcal{L}M_{xy}$, where $M_{x\bar{y}}$ ($M_{xy}$) are mirror reflections at planes perpendicular to $[1\bar{1}0]$ ([110]). 
They enforce the constraint $\textbf{M}_1=\textbf{M}_3=(a,a,b)$, $\textbf{M}_2=\textbf{M}_4=(c,c,d)$, in agreement with the numerical results. The subspace of the conserved symmetries thus allows for patterns in which the magnetization cant away from the field direction without further reducing the symmetry group. 

Analogous to the honeycomb model, a field perpendicular to the lattice plane preserves more symmetries. Here, a $[001]$ field preserves all rotations $R^i,i=0,\dots,3$, as well as $\mathcal{L}R^i$, all mirror symmetries $M_i$ and also $\mathcal{T}M_i$ and $\mathcal{LT}M_i$. Canting is not possible without breaking symmetries. However, such a [001] field is special, as it conserves the flux through the square plaquettes even in the polarized phase. In other words one always has $W_{\square}=-1$ (it's zero field value), no matter the magnitude of the field. 

\subsection{(10,3)a hyperoctagon lattice}
To go beyond the numerically addressed lattices, we apply the above symmetry considerations to the Kitaev model on a variety of three-dimensional lattices \cite{Obrien2016classification}. We study the influence of a [111] field on the magnetization and investigate, if special field directions that do not allow for canting while preserving all symmetries exist. Note that when studying the transformation of the spin operators, we fix the spin axis such that it matches with the lattice coordinate system. The choice of such a global coordinate system is justified, because in real materials, the local spin axes are determined by the octahedral oxygen cage around the Iridium atoms. For the lattices considered here, this allows one to chose the spin axis according to the real space axis \cite{Hermanns2014quantum, Trebst2017}.

\begin{figure}
	\includegraphics[scale=1]{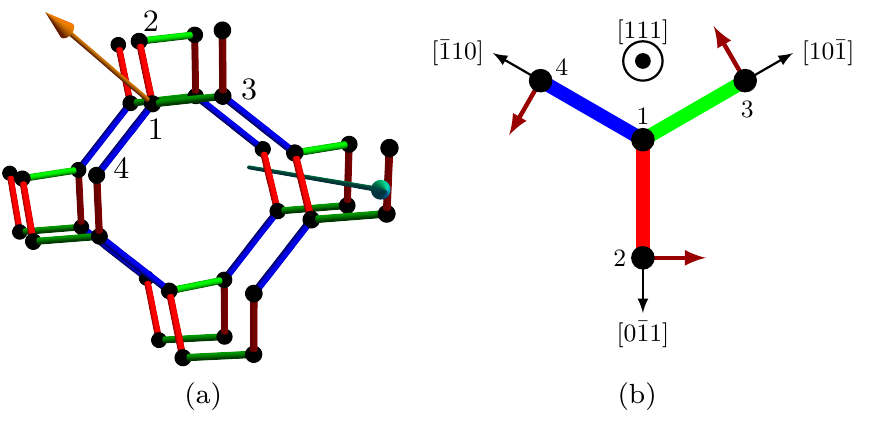}
	\caption{{\bf The (10,3)a hyperoctagon lattice} consists of counter rotating square and octagon spirals as shown in panel (a). It has three four-fold screw axis, one of them is shown in light blue. It is also invariant under a $\pi$ rotation around each lattice bond. The three-fold rotation axis are perpendicular to the plane spanned by a lattice point $i$ and its three neighbors, as visualized here for the [111] axis in orange. (b) When a [111] field is applied, the remaining symmetries are the combination of $\mathcal{T}$ and the rotations around the bonds perpendicular to the field direction, e.g. $[\bar{1}10]$, $[10\bar{1}]$ and $[0\bar{1}1]$ for a [111] field and the three-fold rotation around the field direction. The dark red arrows indicate the symmetry-allowed component of \textbf{M} perpendicular to [111]. Note that \textbf{M}$_1$ is aligned in field direction.}
	\label{fig:103a}
\end{figure}

The lattice (10,3)a is described by the cubic space group $I4_132$ (214).
It hosts a gapless QSL with two Majorana surfaces \cite{Hermanns2014quantum} and consists of counter-rotating spirals formed by squares and octagons that penetrate in the directions of the cubic lattice vectors, as shown in Fig. \ref{fig:103a}.

The space group 214 contains 48 elements that -- as the lattice points sit not a the most generic but at high symmetry positions -- correspond to only 24 symmetry operations for (10,3)a. These operations are (i) the identity, (ii) three four-fold screw-rotations around the square-spirals, i.e. the conventional cubic lattice vectors (9 elements), (iii) four three-fold rotations around $[111]$, $[\bar{1}11]$, $[1\bar{1}1]$ and $[11\bar{1}]$ (8 elements), see Fig. \ref{fig:103a}(a) and (iv) six two-fold rotations around the lattice bonds. The action of these rotations is summarized in Table \ref{table:103a}.

Switching on a [111] field breaks all symmetries except for the three-fold rotation around the field direction and the two-fold rotations around the three bonds perpendicular to the field in combination with $\mathcal{T}$, as visualized in Fig \ref{fig:103a}(b). \textbf{M}$_1$ is aligned in field direction and not related to any other lattice site. For the unit cell positions 2,3,4, the $[111]$ component of \textbf{M} is identical, but the orthogonal components allows the canting pattern sketched in Fig. \ref{fig:103a}(b).

The obvious directions to start looking for fields that preserve more symmetries, are the two remaining rotation axis, pointing in bond direction, e.g. $[110]$ and in the direction of the square spirals, e.g. $[001]$.
However, these two directions also lower the symmetry group substantially and preserve only four and eight symmetries respectively.

\begin{table} 
	\caption{Action of the symmetries operations of the space group $I4_132$ on the Kitaev model on the (10,3)a lattice. The order of the transformations is chosen according to Ref.~\onlinecite{aroyo_bilbao_2006}. The columns 2-5 indicate how the position within a unit cell, defined as in Ref.~\onlinecite{Obrien2016classification}, changes under a transformation. The columns 6-8 describe the behavior of both, bonds and spin operators, e.g. a $-y$ in the $x$-column indicates that $x$-bonds are mapped to $y$-bonds and $\sigma_x \to -\sigma_y$. For (10,3)a, all operations correspond to (screw-)rotations, specified by the axis \textbf{n} and an angle $\alpha$.}
	\label{table:103a}
	\begin{tabular}{r@{\hskip 0.3cm}|@{\hskip 0.3cm} c@{\hskip 0.2cm}c@{\hskip 0.2cm}c@{\hskip 0.2cm}c @{\hskip 0.3cm}|@{\hskip 0.3cm}rrr @{\hskip 0.3cm}|@{\hskip 0.3cm}c@{\hskip 0.3cm}r@{\hskip 0.3cm}r} 
		\toprule 
		&  \multicolumn{4}{c}{sites}  {\hskip 0.5cm}& \multicolumn{3}{c}{bonds} &  \multicolumn{3}{c}{interpretation} \\ 
		& 1 & 2 & 3 & 4 & $x$ & $y$  & $z$ &  \textbf{n} & $\alpha / \pi$ & $T$ \\ 
		\hline 
		1 & 1 & 2 & 3 & 4 & $x$ & $y$ & $z$ & identity & - & -  \\ 
		2 & 4 & 3 & 2 & 1 & $-x$ & $-y$ & $z$ & $001$ & 1 & $\hat{z}/2$   \\ 
		3 & 3 & 4 & 1 & 2 & $-x$ & $y$ & $-z$ &  $010$ & 1 & $\hat{y}/2$   \\ 
		4 & 2 & 1 & 4 & 3 & $x$ & $-y$ & $-z$ &  $100$ & 1 & $\hat{x}/2$  \\ 
		5 & 1 & 3 & 4 & 2 & $y$ & $z$ & $x$ &  $111$ & $2/3$ \\ 
		6 & 2 & 4 & 3 & 1 & $-y$ & $-z$ & $x$ &  $1\bar{1}1$ & $4/3$ \\ 
		7 & 4 & 2 & 1 & 3 & $-y$ & $z$ & $-x$ & $\bar{1}11$ & $4/3$ \\ 
		8 & 3 & 1 & 2 & 4 & $y$ & $-z$ & $-x$ & $11\bar{1}$ & $4/3$ \\ 
		9 & 1 & 4 & 2 & 3 & $z$ & $x$ & $y$ & $111$ & $4/3$  \\ 
		10 & 3 & 2 & 4 & 1 & $-z$ & $-x$ & $y$ & $\bar{1}11$ & $2/3$  \\ 
		11 & 2 & 3 & 1 & 4 & $-z$ & $x$ & $-y$ & $11\bar{1}$ & $2/3$  \\ 
		12 & 4 & 1 & 3 & 2 & $z$ & $-x$ & $-y$ & $1\bar{1}1$ & $2/3$  \\ 
		13 & 4 & 2 & 3 & 1 & $y$ & $x$ & $-z$ & $110$ & 1 \\ 
		14 & 1 & 3 & 2 & 4 & $-y$ & $-x$ & $-z$ & $1\bar{1}0$ & 1 \\ 
		15 & 3 & 1 & 4 & 2 & $-y$ & $x$ & $z$ & $001$ & $-1/2$  &$-\hat{z}/4$ \\ 
		16 & 2 & 4 & 1 & 3 & $y$ & $-x$ & $z$ & $001$ &  $1/2$ & $\hat{z}/4$ \\ 
		17 & 4 & 3 & 1 & 2 & $x$ & $-z$ & $y$ & $100$  &  $-1/2$ & $-\hat{x}/4$ \\ 
		18 & 2 & 1 & 3 & 4 & $-x$ & $z$ & $y$ & $011$ & 1 \\ 
		19 & 1 & 2 & 4 & 3 & $-x$ & $-z$ & $-y$ & $01\bar{1}$ &  1  \\ 
		20 & 3 & 4 & 2 & 1 & $x$ & $z$ & $-y$ & $100$ & $1/2$ & $\hat{x}/4$ \\ 
		21 & 4 & 1 & 2 & 3 & $-z$ & $y$ & $x$ & $010$ & $1/2$ & $\hat{y}/4$ \\ 
		22 & 3 & 2 & 1 & 4 & $z$ & $-y$ & $x$ & $101$ &  $1$ \\ 
		23 & 2 & 3 & 4 & 1 & $z$ & $y$ & $-x$ &  $010$ & $-1/2$ & $-\hat{y}/4$ \\ 
		24 & 1 & 4 & 3 & 2 & $-z$ & $-y$ & $-x$ & $10\bar{1}$ & 1 &\\ 
	\end{tabular}
\end{table}

\subsection{(10,3)b hyperhoneycomb lattice}
\begin{figure} 
	\includegraphics[scale = 1]{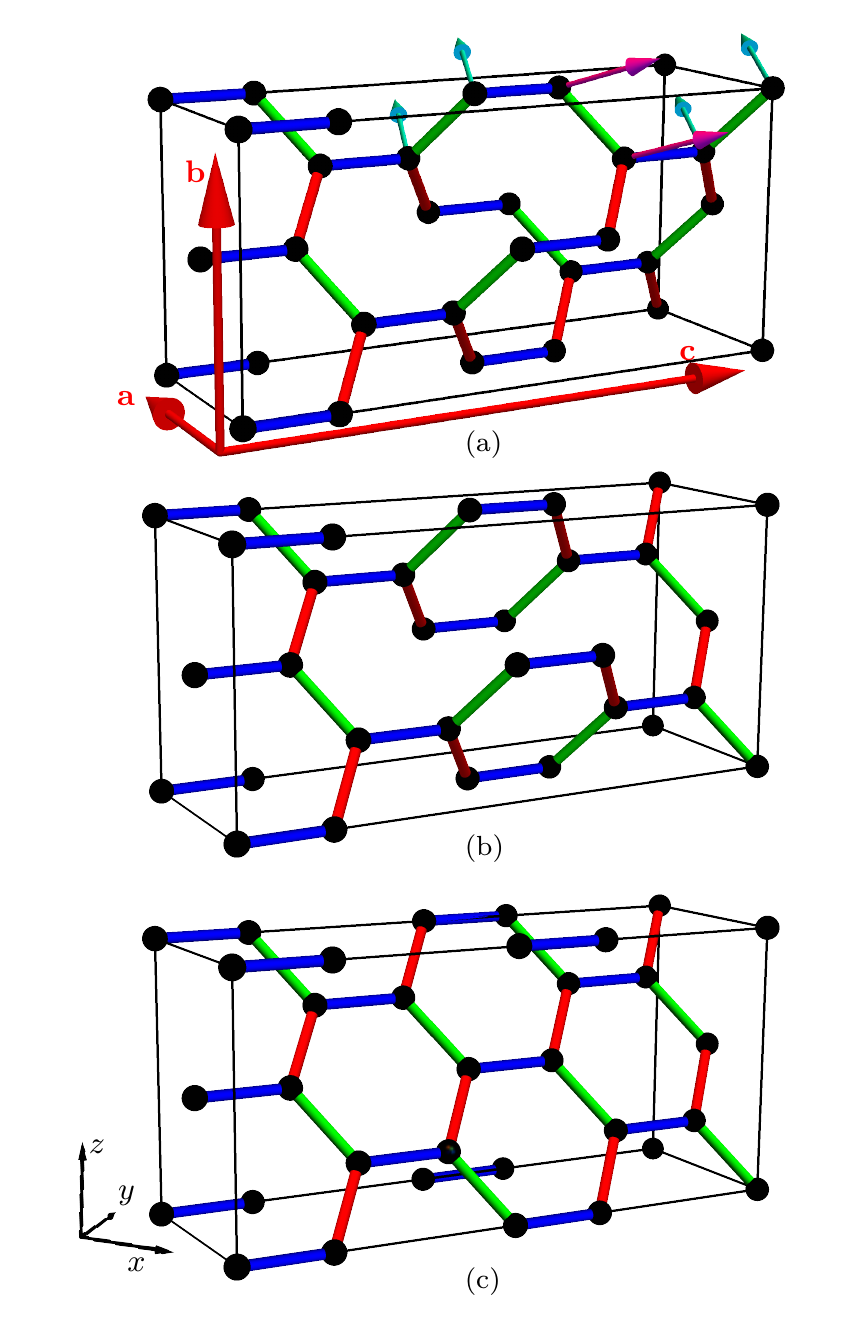}
	\caption{{\bf The hyperhoneycomb lattice and its higher harmonics.} (a) (10,3)b or hyperhoneycomb lattice, consisting of $xy$-zigzag chains that are connected via $z$-bonds aligned in \textbf{c} direction. The light blue and magenta arrows demonstrate the allowed canting pattern in a $[111]$ field. The \textbf{a} components of \textbf{M} are all equal in magnitude but differ in sign for 1,4 (light blue) and 2,3 (magenta).  (b) stripyhoneycomb lattice, consisting of rows of one complete hexagon in \textbf{c} direction, coupled by $z$-bonds. (c) Honeycomb lattice, emerging here as the $N\to\infty$ member of the harmonic honeycomb series, consisting of an arrangement of stripes of $N$ hexagons in \textbf{c} direction coupled by $z$-bonds. All lattices shown here have a twofold rotation symmetry around the \textbf{c} axis, that preserves the unit cell positions.}
	\label{fig:103bstripy}
\end{figure}

The (10,3)b or hyperhoneycomb lattice has recently gained a lot of attention as it is actually realized in the material $\beta$-Li$_2$IrO$_3$. It consists of parallel zigzag chains formed by $x$- and $y$-bonds along two distinct directions, that are coupled by $z$-bonds. For isotropic coupling, the Kitaev model hosts a gapless spin liquid with a nodal line in the Majorana Brillouin zone \cite{Mandal2009exactly, Hermanns2015weyl}. It is described by the orthorombic space group $Fddd$ (70). 
Fig. \ref{fig:103bstripy}(a) visualizes the lattice, where the orthorombic lattice vectors are chosen as $\textbf{a}=(-2,2,0)$, $\textbf{b}=(0,0,4)$, $\textbf{c}=(6,6,0)$. In contrast to (10,3)a, the lattice has one preferred direction -- along \textbf{c} -- and correspondingly, no symmetry operation relates $z$-bonds with $x$- or $y$-bonds. The symmetry group 70 consists of 32 elements. With the same argument as for (10,3)a, these correspond to eight symmetries of the (10,3)b lattice,
that are discussed in Table \ref{table:103b} together with the action on the spin operators. The eight symmetries correspond to (i) the identity, (ii) inversion, with the inversion center being located in the middle of the $x$- or $y$-bonds, (iii) three $\pi$ rotations around the \textbf{a}, \textbf{b} and \textbf{c} vectors, where the rotation axis cuts the $z$-bonds in the middle, and (iv) three glide symmetries, i.e. mirror reflections at the planes spanned by two of the three orthorombic lattice vectors, followed by a translation.

Adding a [111] field conserves three symmetries. Apart from inversion, only rotation around the axis perpendicular to the field and the glide symmetry with the glide plane containing the field direction, both supported by $\mathcal{T}$ remain symmetries of the Kitaev model. Within this subset, canting is possible without breaking additional symmetries. More precisely the constraints are $\textbf{M}_1=\textbf{M}_4$ and $\textbf{M}_2=\textbf{M}_3$ and all magnetization have the same component in \textbf{b} and \textbf{c} directions, but the sign in the \textbf{a} component differs for the lattice sites 1,4 and 2,3, giving rise to the magnetization pattern shown in Fig. \ref{fig:103bstripy}(a).
Similar to the honeycomb case, field directions exists, for which all symmetries listed in Tab. \ref{table:103b} remain a symmetry of the Kitaev model, if  appropriately supported by $\mathcal{T}$. For (10,3)b, there are three such directions, namely the orthorombic lattice vectors. 
Inversion, rotation around the field direction and glide symmetry with the glide plane perpendicular to the field are symmetries whereas the remaining operations have to be supported by $\mathcal{T}$. For a field in \textbf{a} and \textbf{b} direction, the component of all \textbf{M}$_i$ in \textbf{c} direction vanishes and the component in field direction is identical on all lattice sites. The component along the third orthorombic lattice vector on sites 1,4 and 2,3 differs by a sign and therefore, canting is allowed. For a \textbf{c} field, the situation is different. Rotations around the field direction -- symmetry transformation 8 in Table \ref{table:103b} -- do not alter the position within a unit cell but exchange $\sigma_x$ and $\sigma_y$ and change the sign of $\sigma_z$. It follows, that $\textbf{M}$ cannot cant away from the field direction without breaking this symmetry.

\begin{table} 
	\caption{Action of the symmetries operations of the space group $Fddd$ on the Kitaev model on the (10,3)b lattice. The order of the transformations is chosen according to Ref. \cite{aroyo_bilbao_2006}. The unit cell positions in columns 2-5 and the $\textbf{a}_i$ that appear in the translations are taken from  Ref. \cite{Obrien2016classification}. For the glide symmetries in rows 2-4, the column \textbf{n} indicates the direction perpendicular to the glide plane and $T$ the translation that follows the reflection. For the rotations in rows 6-8, \textbf{n} and $\alpha$ give the rotation axis and angle.}
	\label{table:103b}
	\begin{tabular}{r@{\hskip 0.3cm}|@{\hskip 0.3cm}c@{\hskip 0.2cm}c@{\hskip 0.2cm}c@{\hskip 0.2cm}c@{\hskip 0.3cm}|r@{\hskip 0.3cm}r@{\hskip 0.3cm}r@{\hskip 0.3cm}|@{\hskip 0.3cm}c@{\hskip 0.3cm}c@{\hskip 0.3cm}c} 
		\toprule
		& \multicolumn{4}{c}{sites} & \multicolumn{3}{c}{bonds} &  \multicolumn{3}{c}{interpretation} \\ 
		& 1 & 2 & 3 & 4 & $x$ & $y$  & $z$ &  \textbf{n} & $\alpha / \pi$ & $T$ \\ 
		\hline 
		1 & 1 & 2 & 3 & 4 & $x$ & $y$ & $z$ & identity & - & - \\ 
		2 & 2 & 1 & 4 & 3 & $-y$ & $-x$ & $-z$ & \textbf{a} & - & $(\textbf{a}_3\!-\!\textbf{a}_2)/2$   \\ 
		3 & 2 & 1 & 4 & 3 & $-x$ & $-y$ & $z$ &  \textbf{b} & - & $\textbf{a}_3/2$   \\ 
		4 & 4 & 3 & 2 & 1 & $y$ & $x$ & $-z$ &  \textbf{c} & - & $\textbf{a}_1/2$  \\ 
		5 & 4 & 3 & 2 & 1 & $x$ & $y$ & $z$ &  inversion & - & - \\ 
		6 & 3 & 4 & 1 & 2 & $-y$ & $-x$ & $-z$ &  \textbf{a} & 1 & - \\ 
		7 & 3 & 4 & 1 & 2 & $-x$ & $-y$ & $z$ & \textbf{b} & 1 & -\\ 
		8 & 1 & 2 & 3 & 4 & $y$ & $x$ & $-z$ & \textbf{c} & 1 & - \\ 
	\end{tabular}
\end{table}

\subsection{Stripy honeycomb and harmonic honeycomb series}
The arguments from the preceding section remain valid for another three-dimensional lattice that is realized in a material, $\gamma$-Li$_2$IrO$_3$, the stripyhoneycomb lattice, sketched in Fig. \ref{fig:103bstripy}(b) \cite{Kruger2020, Modic2014realization}. Its space group is $Cmmm$ (66). It consists of two rows of complete honeycombs along two distinct directions, that are coupled by $z$-bonds. Again, the 16 elements of $Cmmm$ reduce to eight distinct transformations of the lattice. They have the exact same interpretation as for the (10,3)b lattice, reflections at the planes perpendicular to the \textbf{a}, \textbf{b} and \textbf{c} vectors, followed by a translation, $\pi$ rotations around these vectors and inversion symmetry. As for (10,3)b, fields parallel to the \textbf{a},\textbf{b},\textbf{c} directions conserve eight symmetries. For fields in \textbf{a} and \textbf{b} direction, the same canting pattern as for (10,3)b is possible. A \textbf{c} field does not allow any deviation of \textbf{M} from the field direction as this violates the rotation symmetry around the $z$-bonds, that maps each unit cell position to itself. 
In fact, (10,3)b and stripyhoneycomb are only the first representatives of a whole family of lattices, called the harmonic honeycomb lattices \cite{Kimchi2014three, Modic2014realization}, consisting of rows of $N$ complete honeycombs in \textbf{c} direction that are connected by $z$-bonds pointing also along \textbf{c}. Two successive honeycomb rows in \textbf{c} directions are rotated against each other. For $N=0,1,\infty$, we regain the (10,3)b, stripyhoneycomb and honeycomb lattice, as depicted in Fig. \ref{fig:103bstripy}. As the \textbf{c} rotation is preserved for all members of the harmonic honeycomb series, we expect a field in this direction to generically forbid canting. Indeed, we saw for the honeycomb lattice, the $N=\infty$ case, that a combination of mirror symmetry and inversion prevents canting. When interpreting the honeycomb lattice as a true 3d object, this combination correspond indeed to the \textbf{c} axis rotation.

\end{document}